\documentclass{emulateapj}

\slugcomment{To Appear in  {\it The Astrophysical Journal}.}

\begin{document}

\title{A New Sample of Candidate Intermediate-Mass Black Holes Selected by X-ray Variability\footnote
{
Based on observations obtained with {\it XMM-Newton}, an ESA science mission with
instruments and contributions directly funded by ESA Member States and NASA
}
}

\author{Naoya Kamizasa, Yuichi Terashima, and Hisamitsu Awaki}
\affil{Department of Physics, Ehime University, Matsuyama, Ehime 790-8577, Japan}

\begin{abstract}

We present the results of X-ray variability and spectral analysis of 
a sample of 15 new candidates for active galactic nuclei
with relatively low-mass black holes (BHs).
They are selected from the Second {\it XMM-Newton} Serendipitous
Source Catalogue based on strong variability quantified by normalized
excess variances. Their BH masses
are estimated to be $(1.1-6.6)\times10^6$ $M_\odot$ by using
a correlation between excess variance and BH mass.
Seven sources have estimated BH masses smaller than $2\times10^6$ $M_\odot$,
which are in the range for intermediate-mass black holes.
Eddington ratios of sources with known
redshifts range from 0.07 to 0.46 and the mean Eddington ratio is 0.24.
These results imply that some of our sources are growing supermassive black holes,
which are expected to have relatively low masses with high Eddington ratios. X-ray photon
indices of the 15 sources are in the range of $\approx0.57-2.57$, and 5 among
them have steep ($>2$) photon indices, which are the range for
narrow-line Seyfert 1s. Soft X-ray excess is seen in 12 sources, and is
expressed by a blackbody model with $kT\approx83-294$ eV. We derive
a correlation between X-ray photon indices and Eddington ratios,
and find that the X-ray photon indices of about a half of our sources are
flatter than the positive correlation suggested previously. 

\end{abstract}


\section{INTRODUCTION}

Active galactic nuclei (AGNs) contain supermassive black holes (SMBHs) with masses in the range of $\sim 10^{6}-10^{9}M_\odot$, 
and their evolution is believed to be closely related to their host galaxies 
(e.g., Magorrian et al 1998; Marconi \& Hunt 2003; H\"aring \& Rix 2004). 
It is, however, still not understood why such relation exists. 
One way to approach this issue is to search for and study SMBHs in their growing phase. 
SMBHs are considered to grow up by accretion and/or merging. 
Marconi et al. (2004) suggested that the growth of local SMBHs (redshift $z<3$) 
is mainly governed by accretion, and that accretion is likely to play an important role in the growth process of SMBHs. 
If accretion is indeed a major process of black hole (BH) growth, growing SMBHs are expected to have a combination of 
relatively low masses and high accretion rates.
In fact, recent studies found AGNs harboring candidate intermediate-mass black holes (IMBHs) with 
masses $\sim10^{4}-10^{6}$ $M_\odot$ 
by using optical emission line widths to estimate BH mass (Filippenko \& Ho 2003; Barth et al. 2004; 
Greene \& Ho 2004, 2007a; Dong et al. 2007). 
Greene \& Ho (2007a) selected 174 AGNs with candidate IMBHs from the forth data release of the Sloan Digital Sky Survey
(SDSS), 
and showed that these AGNs are radiating at high fractions of their Eddington limits with a median Eddington ratio of 0.4.
This class with relatively low-mass BHs with high accretion rates is a candidate of growing BHs
and is crucial to study physical processes of accretion and mass growth.

High energy emission, which is from the vicinity of central BHs, 
is an essential probe of accretion processes and X-ray observations of AGNs with IMBHs are of great interest.
Among studies with X-rays (Greene \& Ho 2007b, Desroches et al. 2009, Dewangan et al. 2008, Miniutti et al. 2009),
relatively good quality X-ray spectra were obtained with {\it XMM-Newton} for AGNs with candidate IMBHs 
selected from the sample of  Greene \& Ho (2004). 
At least four objects show clear signature of the presence of soft excess emission 
represented by a multicolor disk blackbody model (Mitsuda et al. 1984) with an inner temperature 0.15--0.2 keV (Dewangan et al. 2008), 
which is similar to that observed in
more massive SMBHs (Gierli\'nski \& Done 2004; Crummy et al. 2006; Bianchi et al. 2009).
Photon indices for their sample are in the range of 1.6--2.4, which are also in the range 
seen in SMBHs (Lu \& Yu 1999; Shemmer et al. 2006, 2008; Risaliti et al. 2009).
These studies are limited to relatively X-ray bright objects selected by the optical band, 
and a new method using X-rays to select IMBH candidates would be useful to compile X-ray bright IMBHs
for further  studies of X-ray properties and accretion physics.

X-ray variability is a well-known property of AGNs, and its timescale is related to a size of X-ray emitting region. 
BH masses can be estimated from X-ray variability by assuming that the size of the emitting region is proportional to BH mass. 
For instance, MacHardy et al. (2006) estimated BH masses by using power spectral density (PSD), 
which can be used to identify characteristic timescales. On the other hand, the variability amplitude was often quantified 
by the normalized excess variance (NXS). 
Since NXS represents an integration of the PSD in a certain frequency range normalized 
by the mean count rate squared (Vaughan et al. 2003), NXS can be also used to estimate BH masses. 
An anti-correlation between NXS and BH masses was found by various authors (Lu \& Yu 2001; Bian \& Zhao 2003; 
Papadakis 2004; O'Neill et al. 2005; Miniutti et al. 2009; Niko{\l}ajuk et al. 2009; Zhou et al. 2010). 
Indeed, optically selected AGNs with IMBHs show large amplitude variability on short timescales (Dewangan et al. 2008).
These results indicate that highly variable AGNs are strong candidates for hosts of IMBHs and that X-ray variability can be used
to find IMBHs.
One advantage of a method utilizing X-ray variability to search for IMBHs
compared with a method based on optical spectra is that IMBHs can be found even if they are moderately obscured (up to $N_{\rm H}<{\rm several}\times10^{23}$~cm$^{-2}$).

\begin{deluxetable*}{lcccccc}
\tabletypesize{\scriptsize}
\tablecaption{Observation Log\label{table:sample}}
\tablewidth{0pt}
\tablehead{\colhead{Name} & \colhead{Alternative name} & \colhead{ObsID} & \colhead{Start Date} & \colhead{Redshift} & \colhead{Mean Rate$^{\rm a}$} & \colhead{Exposure$^{\rm b}$}\\
 & & & & & \colhead{(counts s$^{-1}$)} & \colhead{(ks)}}
\startdata
2XMM J002133.3$-$150751 & SDSS J002133.39$-$150752.3 & 0203460101 & 2004 Dec 22 & 0.135$^{\rm c}$ & 0.041 & 37.5\\
2XMM J011356.4$-$144239 & 2MASX J01135640$-$1442401 & 0147920101 & 2003 Jun 15 & 0.054$^{\rm d}$ & 0.088 & 18.3\\
2XMM J013612.5+154957 & CXO J013612.5+154957 & 0154350101& 2002 Feb 01 & \nodata & 0.044 & 25.8\\
2XMM J015234.8$-$134735 & CXO J015234.8$-$134736 & 0109540101 & 2002 Dec 24 & 0.17$^{\rm d}$ & 0.050 & 43.5\\
2XMM J023213.4$-$072945 & SDSS J023213.43$-$072946.4 & 0200730401 & 2004 Jan 07 & 0.159$^{\rm c}$ & 0.041 & 35.0\\
2XMMi J032459.9$-$025612 & 2MASX J03245992-0256122 & 0405240201 & 2006 Aug 06 & $\cdots$ & 0.118 & 11.1\\
2XMM J120143.6$-$184857 & CXO J120143.6$-$184857 & 0085220101 & 2002 Jan 18 & \nodata & 0.058 & 14.9\\
2XMM J123103.2+110648$^{\rm e}$ & SDSS J123103.23+110648.6 & 0306630101 & 2005 Dec 13 & 0.128$^{\rm c}$ & 0.091 & 54.6\\
2XMM J123316.6+000512 & SDSS J123316.65+000511.5 & 0203170301 & 2004 Dec 25 & 0.196$^{\rm c}$ & 0.035 & 64.0\\
2XMM J130543.9+181355 & SDSS J130543.96+181356.0 & 0017940101 & 2001 Jan 03 & 0.171$^{\rm c}$ & 0.042 & 40.0\\
2XMM J132419.0+300042 & SDSS J132418.98+300042.0 & 0025740201 & 2001 Dec 12 & 0.117$^{\rm c}$ & 0.084 & 30.8\\
2XMM J134736.4+173404 & SDSS J134736.39+173404.6 & 0144570101 & 2003 Jun 24 & 0.0447$^{\rm f}$ & 0.285 & 37.7\\
2XMM J200824.5$-$444009 & 2MASX J20082452$-$4440095 & 0200360201 & 2004 Apr 11 & 0.0581$^{\rm d}$ & 0.639 & 18.6\\
2XMM J213152.8$-$425130 & \nodata & 0200780301 & 2004 Oct 28 & \nodata & 0.038 & 26.3\\
2XMMi J233430.3+392101 & 2MASX J23343041+3920596 & 0305570101 & 2006 Jan 02 & \nodata & 0.405 & 27.2\\
2XMM J235509.6+060041 & \nodata & 0206060101 & 2004 Jun 14 & \nodata & 0.088 & 15.1\\
\enddata
\tablenotetext{a}{Count rate in 0.2$-$12 keV taken from 2XMMi-DR3.}
\tablenotetext{b}{Effective exposure time for EPIC-pn.}
\tablenotetext{c}{Photometric redshift based on photometric data of SDSS.}
\tablenotetext{d}{Redshift taken from NED.}
\tablenotetext{e}{Analysis of this source will be given in a forthcoming paper.}
\tablenotetext{f}{Redshift based on SDSS spectrum.}
\end{deluxetable*}

The {\it XMM-Newton} satellite carries X-ray telescopes with the largest effective area, and is suitable to detect rapid variability, 
which is a signature of the presence of a relatively low-mass SMBH. 
Furthermore, a serendipitous survey carried out with its large field of view is effective to search for rare populations like IMBHs. 
The Second {\it XMM-Newton} Serendipitous Source Catalogue (2XMMi-DR3) is the largest X-ray source catalogue, 
which contains various types of X-ray sources (262902 unique sources) drawn from a wide-area serendipitous survey 
(Watson et al. 2009).
We select highly variable AGNs from 2XMMi-DR3 to search for relatively low-mass SMBHs 
and successfully find 16 candidates for new AGNs hosting relatively low-mass SMBHs as well as previously known IMBHs.
In this paper, we describe the results of X-ray variability and spectral analysis of 15 objects among them,
excluding 2XMM J123103.2+110648.
The latter object  shows a peculiar X-ray spectrum explained by thermal emission only
and detailed results are presented in Terashima et al. (2012).
This paper is organized as follows.
The selection method and data reduction are described in Section 2.
We present the results of variability and spectral analysis in Section 3, 
and discuss the results in Section 4. The conclusions are given in Section 5. In this paper, we assume the cosmological parameters $H_0=70$ km s$^{-1}$ Mpc$^{-1}$, $\Omega_{\rm m}=0.3$, and $\Omega_\Lambda=0.7$.

\section{SAMPLE SELECTION AND DATA REDUCTION}

Our sample was selected from 2XMMi-DR3 produced by the {\it XMM} Survey Science Centre (Watson et al. 2009). 
In order to search for highly variable objects, we utilized the $\chi^2$ probabilities that the source is constant listed in 2XMMi-DR3. 
These probabilities were calculated from the time series in 0.2$-$12 keV 
by using the Science Analysis System (SAS) task {\tt ekstest} excluding high background flaring times.
We selected sources satisfying the following conditions:
(1) the probability calculated by using EPIC-pn data $<10^{-5}$, 
(2) the count rate for EPIC-pn in 0.2$-$12 keV $\ge0.03$ counts s$^{-1}$, and 
(3) Galactic latitude $|b|>10^\circ$, where all of these parameters are listed in 2XMMi-DR3. 
1100 sources fulfilled these criteria. We applied further filtering conditions to construct our sample.
We discarded sources in the fields of star forming regions, Small Magellanic Clouds, or Large Magellanic Clouds, 
since these regions contain many X-ray sources, most of which are unlikely to be AGN. 
Sources with object types Galaxy, X-ray source, or unidentified
shown in the NASA/IPAC Extragalactic Database (NED) or the SIMBAD were regarded as AGN candidates. 
In addition, we examined X-ray images, light curves, and spectra, and excluded the following sources:
(1) sources in the fields of extended objects such as supernovae remnants or clusters of galaxies, 
(2) sources with effective exposure time $\le10$ ks, 
(3) sources showing light curves like stellar flares, i.e., abrupt rise of flux followed by exponential decay,
and
(4) sources showing spectra dominated by thermal plasma emission.
After all the screening, 59 sources were finally selected.
16 among them are not classified as AGNs in the literature published so far and are candidates for new AGNs.
We examined optical images of the Digitized Sky Survey to search for optical counterparts.
For 15 sources among the 16 sources, there is only one optical source within the error circle of the X-ray position.
There is one optical source within the error circle of X-ray position of J1347+1734,
and another source is located just outside of the circle.
There is no optical source inside the error circle of the position J1204$-$1848.
The optical sizes are much smaller than the error circle of the X-ray sources in all the possible optical counterpart,
and the possibility that the X-ray sources are off-center cannot be ruled out.
This paper presents X-ray variability and spectral analysis of 15 objects among the 16 new AGN candidates, 
excluding 2XMM J123103.2+110648 presented in Terashima et al. (2012).
In Table \ref{table:sample}, we list the source names, start date of observations, EPIC-pn count rates in the 0.2$-$12 keV band, 
and effective exposure times. Spectroscopic or photometric redshifts are also shown in Table \ref{table:sample} when available. 

We analyzed X-ray light curves and spectra of the selected sources obtained by EPIC-pn. 
The observation data files were reprocessed to produce the calibrated event files by using the 
SAS version 9.0.0. and calibration data as of 2010 March.
The X-ray event patterns 0$-$4 were selected. We extracted source light curves and spectra from circular regions 
centered at the target with a radius of 20$''$$-$30$''$.
Time intervals with high background rates seen in light curves of an off-source region in 10$-$12 keV were discarded.
Background spectra were created from a rectangular region free of sources on the same CCD chip.
The {\tt arfgen} and {\tt rmfgen} tasks in SAS were used to generate the ancillary response files and detector response matrices, respectively. 
In order to use the $\chi^2$ fits, the spectra were binned so that each bin contains at least 25 counts.
For spectral analysis, we used XSPEC version 12.6.0.
The errors on the spectral parameters are quoted at a 90\% confidence level for one parameter of interest.

\section{ANALYSIS}
\subsection{Light Curves}

We prepared  light curves in the 0.5$-$10 keV band with 512 s bin, and quantified variability 
with the NXS. 
Fig.\,\ref{fig:lc} shows source and background light curves in the 0.5$-$10 keV band with 512 s bin. 
The background rates are normalized to the area of the source light curves.
When light curves were made, the redshift-corrected energy band was used for sources with a redshift greater than 0.1. 
NXS is the variance after correcting for measurement errors 
(e.g., Nandra et al. 1997; Turner et al. 1999; Vaughan et al. 2003) and is defined by
\begin{equation}
\sigma^2_{\rm NXS}=\frac{1}{{\bar x}^2}\left[\frac{1}{N-1}\sum^N_{i=1}(x_i-{\bar x})^2-\frac{1}{N}\sum^N_{i=1}\sigma^2_i\right],
\end{equation}
where $N$ and ${\bar x}$ are the number of data points and the mean count rate, respectively. 
$x_i$ is the count rate in the $i$-th bin, and its error is $\sigma_i$. 
The error of NXS was estimated by using the expression of Vaughan et al. (2003). 
Previous studies of variability often used light curves in the 2$-$10 keV band with 256 s bin to calculate NXS. 
Light curves of some sources, however, contain bins of 0 counts s$^{-1}$, and NXS cannot be properly calculated. 
We, therefore, calculated NXS by using light curves in the 0.5$-$10 keV band with 512 s bin, 
and only few bins of 0 counts s$^{-1}$ are seen in the light curves. Light curves of J0021$-$1507 and J1305+1813 still contain one 
and three bins of 0 counts s$^{-1}$ among 80 and 88 bins, respectively. 
In these cases, since the fractions of such bins were small and their influence on the NXS calculation is small, 
we excluded such bins from the calculation of NXS.

\begin{deluxetable*}{lcccccccccccc}
\tabletypesize{\scriptsize}
\tablecaption{Spectral Parameters for the Best-fit Models\label{table:spf}}
\tablewidth{0pt}
\tablehead{\colhead{Name} & \colhead{$N_{\rm H,Gal}$$^{\rm a}$} & \colhead{$N_{\rm H}$$^{\rm b}$} & \colhead{$\Gamma$$^{\rm c}$} & \colhead{$n_{\rm PL}$$^{\rm d}$} & \colhead{$kT$$^{\rm e}$} & \colhead{$n_{\rm BB}$$^{\rm f}$} & \colhead{$E$$^{\rm g}$} & \colhead{$\tau$$^{\rm h}$} & \colhead{$\chi^2_\nu$ (dof)} \\
 & \colhead{($10^{20}$ ${\rm cm}^{-2}$)} & \colhead{($10^{20}$ ${\rm cm}^{-2}$)} & & \colhead{($10^{-6}$)} & \colhead{(keV)} & \colhead{($10^{-7}$)} & \colhead{(keV)} & &  }
\startdata
J0021$-$1507 & 2.07 (f) & $<6.2$ & $1.55^{+0.20}_{-0.13}$ & $21.6^{+5.7}_{-7.2}$ & $0.167^{+0.087}_{-0.098}$ & $2.4^{+3.7}_{-2.3}$ & $0.71^{+0.02}_{-0.03}$ & $2.2^{+1.2}_{-0.9}$ & 0.97 (25) \\
J0136+1549 & 4.50 (f) & $<20.2$ & $1.26^{+0.66}_{-0.73}$ & $7.7^{+6.1}_{-4.3}$ & $0.140^{+0.019}_{-0.020}$ & $10.8^{+1.9}_{-2.5}$ & $0.75^{+0.06}_{-0.04}$ & $1.4\pm0.8$ & 1.03 (16) \\
J0152$-$1347 & 1.40 (f) & $<34$ & $1.49^{+0.15}_{-0.14}$ & $15.6\pm2.7$ & $0.090^{+0.012}_{-0.009}$ & $12.3^{+3.6}_{-3.3}$ & \nodata & \nodata & 1.32 (25) \\
J0232$-$0729 & 3.14 (f) & $<6.6$ & $2.21^{+0.13}_{-0.12}$ & $24.3\pm1.8$ & \nodata & \nodata & \nodata & \nodata & 0.56 (22) \\
J0324$-$0256 & 3.73 (f) & $<2.0$ & $1.89^{+0.11}_{-0.10}$ & $50.5\pm3.3$ & \nodata & \nodata & \nodata & \nodata & 0.96 (31) \\
J1201$-$1848 & 3.22 (f) & $<13$ & $2.20^{+0.34}_{-0.28}$ & $11.8^{+7.1}_{-4.2}$ & $0.127^{+0.022}_{-0.019}$ & $7.7^{+3.8}_{-5.3}$ & \nodata & \nodata & 0.80 (17) \\
J1233+0005 & 1.89 (f) & $<20$ & $2.02^{+0.16}_{-0.11}$ & $17.6^{+2.3}_{-2.7}$ & $0.089^{+0.027}_{-0.018}$ & $4.6^{+6.4}_{-2.3}$ & \nodata & \nodata & 0.65 (44)\\
J1305+1813 & 1.97 (f) & $<42$  & $2.00^{+0.61}_{-0.36}$ & $6.5^{+8.4}_{-4.9}$ & $0.163^{+0.039}_{-0.020}$ & $7.0^{+2.0}_{-4.5}$ & \nodata & \nodata & 1.28 (16) \\
J1324+3000 & 1.25 (f) & $<11$ & $0.57^{+0.22}_{-0.15}$ & $7.5^{+6.7}_{-4.6}$ & $0.294^{+0.028}_{-0.027}$ & $9.5^{+1.8}_{-1.9}$ & \nodata & \nodata & 1.09 (33) \\
J1347+1734 & 1.78 (f) & $<2.1$ & $2.57^{+0.07}_{-0.06}$ & $135^{+13}_{-11}$ & $0.163^{+0.033}_{-0.019}$ & $13.6^{+8.0}_{-7.1}$ & \nodata & \nodata & 1.01 (138) \\
J2008$-$4440 & 3.24 (f) & $<3.4$ & $1.83\pm0.06$ & $214^{+14}_{-12}$ & $0.103\pm0.005$ & $98.1^{+8.5}_{-8.4}$ & \nodata & \nodata & 0.97 (191) \\
J2131$-$4251 & 2.56 (f) & $<6.8$ & $1.76^{+0.17}_{-0.15}$ & $13.4^{+2.0}_{-2.5}$ & $0.083^{+0.034}_{-0.025}$ & $2.8^{+9.7}_{-1.6}$ & \nodata & \nodata & 1.04 (27) \\
J2334+3921 & 8.33 (f) & $<3.1$ & $2.38\pm0.05$ & $157^{+12}_{-13}$ & $0.138\pm0.007$ & $45.8^{+8.6}_{-9.0}$ & \nodata & \nodata & 1.07 (203) \\
J2355+0600 & 4.77 (f) & $<17$ & $2.31^{+0.20}_{-0.16}$ & $28.0^{+8.1}_{-6.3}$ & $0.119\pm0.020$ & $9.6^{+4.4}_{-6.3}$ & \nodata & \nodata & 1.15 (35) \\
\enddata 
\tablecomments{(f) indicates fixed parameter.}
\tablenotetext{a}{Galactic column density derived from 21 cm measurement (Kalberla et al. 2005).}
\tablenotetext{b}{Column density of hydrogen at the redshift of the source.}
\tablenotetext{c}{Photon index of power law.}
\tablenotetext{d}{Normalization of power law in units of photons keV$^{-1}$ cm$^{-2}$ s$^{-1}$ at 1 keV.}
\tablenotetext{e}{Temperature of blackbody.}
\tablenotetext{f}{Normalization of the blackbody, $L_{39}/D^2_{10}$, where $L_{39}$ is the source luminosity in units of $10^{39}$ erg s$^{-1}$, and $D_{10}$ is the distance to the source in units of 10 kpc.}
\tablenotetext{g}{Threshold energy of absorption edge.}
\tablenotetext{h}{Absorption depth at the threshold energy.}
\end{deluxetable*}

\begin{deluxetable*}{cccccccccc}
\tabletypesize{\scriptsize}
\tablecaption{Spectral Parameters for J0113$-$1442\label{table:spf2}}
\tablewidth{0pt}
\tablehead{\colhead{$N_{\rm H,Gal}$$^{\rm a}$} & \colhead{$N_{\rm H}$$^{\rm b}$} & \colhead{C.F.$^{\rm c}$}
& \colhead{$\Gamma$$^{\rm d}$} & \colhead{$n_{\rm PL}$$^{\rm e}$}
& \colhead{$E$$^{\rm f}$} & \colhead{$\tau$$^{\rm g}$}
& \colhead{$kT$$^{\rm h}$} & \colhead{$n_{\rm BB}$$^{\rm i}$}
& \colhead{$\chi^2_\nu$ (dof)}\\
\colhead{($10^{20}$ ${\rm cm}^{-2}$)} & \colhead{($10^{22}$ ${\rm cm}^{-2}$)}& & & \colhead{($10^{-4}$)} & \colhead{(keV)} & & \colhead{(keV)} & \colhead{($10^{-6}$)} &  }
\startdata
1.72 (f) & $6.6^{+2.0}_{-1.4}$ & $0.90^{+0.04}_{-0.08}$ & $1.98\pm0.27$ & $2.3^{+1.3}_{-0.8}$ & $0.70^{+0.05}_{-0.07}$ & $1.37^{+0.88}_{-0.66}$ & \nodata & \nodata & 1.16 (21)\\
1.72 (f) & $4.0\pm1.8$ & $0.82^{+0.09}_{-0.21}$ & $1.34^{+0.47}_{-0.48}$ & $0.72^{+0.88}_{-0.39}$ & \nodata & \nodata & $0.077\pm0.026$ & $4^{+17}_{-2}$ & 1.13 (21)
\enddata
\tablecomments{(f) indicates fixed parameter.}
\tablenotetext{a}{Galactic column density derived from 21 cm measurement (Kalberla et al. 2005).}
\tablenotetext{b}{Column density of hydrogen at the redshift of the source.}
\tablenotetext{c}{Covering fraction.}
\tablenotetext{d}{Photon index of power law.}
\tablenotetext{e}{Normalization of the power law absorbed by $N_{\rm H}$ in units of photons keV$^{-1}$ cm$^{-2}$ s$^{-1}$ at 1 keV.}
\tablenotetext{f}{Threshold energy of absorption edge.}
\tablenotetext{g}{Absorption depth at the threshold energy.}
\tablenotetext{h}{Temperature of blackbody.}
\tablenotetext{i}{Normalization of the blackbody, $L_{39}/D^2_{10}$, where $L_{39}$ is the source luminosity in units of $10^{39}$ erg s$^{-1}$, and $D_{10}$ is the distance to the source in units of 10 kpc.}
\end{deluxetable*}

\subsection{Spectra}

We examined several models to fit spectra of the 15 strongly variable AGN candidates. 
All the model components were absorbed by the Galactic \ion{H}{1} column density using the {\tt phabs} model in XSPEC. 
We fixed the Galactic \ion{H}{1} column density at the value derived from 21 cm observations (Kalberla et al. 2005).
In all spectral fits, if the redshift is known, 
we used models taking account of the source redshift, to fit the observed spectra.

Spectra of the two sources J0232$-$0729 and J0324$-$0256 were explained by a simple power-law model 
without intrinsic absorption. 
A combination of power law and blackbody reproduced spectra of the 10 sources J0152$-$1347, J1201$-$1848, J1233+0005, J1305+1813, J1324+3000,  J1347+1734, J2008$-$4440, J2131$-$4251, J2334+3921, and J2355+0600. 
The results of these spectral fits are summarized in Table \ref{table:spf}. 
These 12 objects do not require additional absorption intrinsic to the source. We derived upper limits on
their intrinsic absorption column densities by adding an additional absorption. The results are also shown in Table \ref{table:spf}. 
Among these objects, the best-fit photon index for one object (J1324+3000) is rather flat (photon index $\Gamma \approx 0.57$),
which might be due to intrinsic absorption.
Then we applied intrinsic absorption to both the power law and blackbody component.
The spectrum, however, did not require additional intrinsic absorption.
Thus we adopt the result of the fit with the power law plus blackbody model in the following discussions.

These simple continuum models failed to fit the spectra of J0021$-$1507, J0113$-$1442, and J0136+1549
($\chi^2_\nu$ (dof) = 1.73 (27), 1.54 (23), and 1.51 (18), respectively).
When the spectra of J0021$-$1507 and J0136+1549 were fitted by a combination of power law and blackbody, 
an edge-like feature was seen at $\sim0.7$ keV in the residuals. 
Thus, we added an absorption edge model to represent this feature, and obtained better fits
($\Delta\chi^2=24.0$ and 10.6 for J0021$-$1507 and J0136+1549, respectively).
The parameters for this model are shown in Table \ref{table:spf}.

In the spectral fit to J0113$-$1442 with a combination of power law and blackbody,
the photon index obtained was extremely flat ($\sim0.4$).
In order to test whether this flatness is due to absorption,
we applied the intrinsic absorption to the power law and blackbody model.
This, however, resulted in an unacceptable fit ($\chi^2_\nu~({\rm dof})=1.60~(22)$).
We also tried to fit the spectrum using a partially covered power-law model,
\begin{equation}
{\tt phabs\times zpcfabs\times zpowerlaw},
\end{equation}
where {\tt phabs}, {\tt zpcfabs}, and {\tt zpowerlaw} represent models of Galactic absorption, partial covering absorption, 
and power law, respectively.
Although we obtained a steeper photon index $\Gamma\approx2.1$,
the fit was unacceptable ($\chi^2_\nu~({\rm dof})=1.67~(23)$).
In this fit, edge-like residuals were seen at $\sim 0.7$ keV.
Then we added an absorption edge model ``{\tt zedge}",
\begin{equation}
{\tt phabs\times zpcfabs\times zedge\times zpowerlaw},
\end{equation}
and obtained a better fit ($\chi^2_\nu~{\rm (dof)}=1.16~(21)$) with a photon index $\Gamma\approx1.98$.
The spectral parameters of this model are shown in the first row of Table \ref{table:spf2}. 
We also examined the following model,
\begin{equation}
{\tt phabs\times zpcfabs\times(zpowerlaw+zbbody)}.
\end{equation}
We again obtained a good fit ($\chi^2_\nu~{\rm (dof)}=1.13~(21)$) as shown in the second row of Table \ref{table:spf2}.
The spectrum of J0113$-$1442 was almost equally well reproduced by the models (3) and (4).

Spectra of all the sources with the best-fit model are shown in Fig.$\;$\ref{fig:sp}.
The model (3) is shown as the best-fit model for J0113$-$1442.
Observed fluxes and absorption-corrected luminosities in the 2$-$10 keV band  were calculated by using the best-fit model, where luminosities are derived only for objects with known redshifts.
We obtained luminosities in the range of $10^{41}-10^{43}$ erg s$^{-1}$ as summarized in Table \ref{table:REDD}.


\begin{figure*}
\includegraphics[width=7.1cm,height=8.75cm,clip,angle=270]{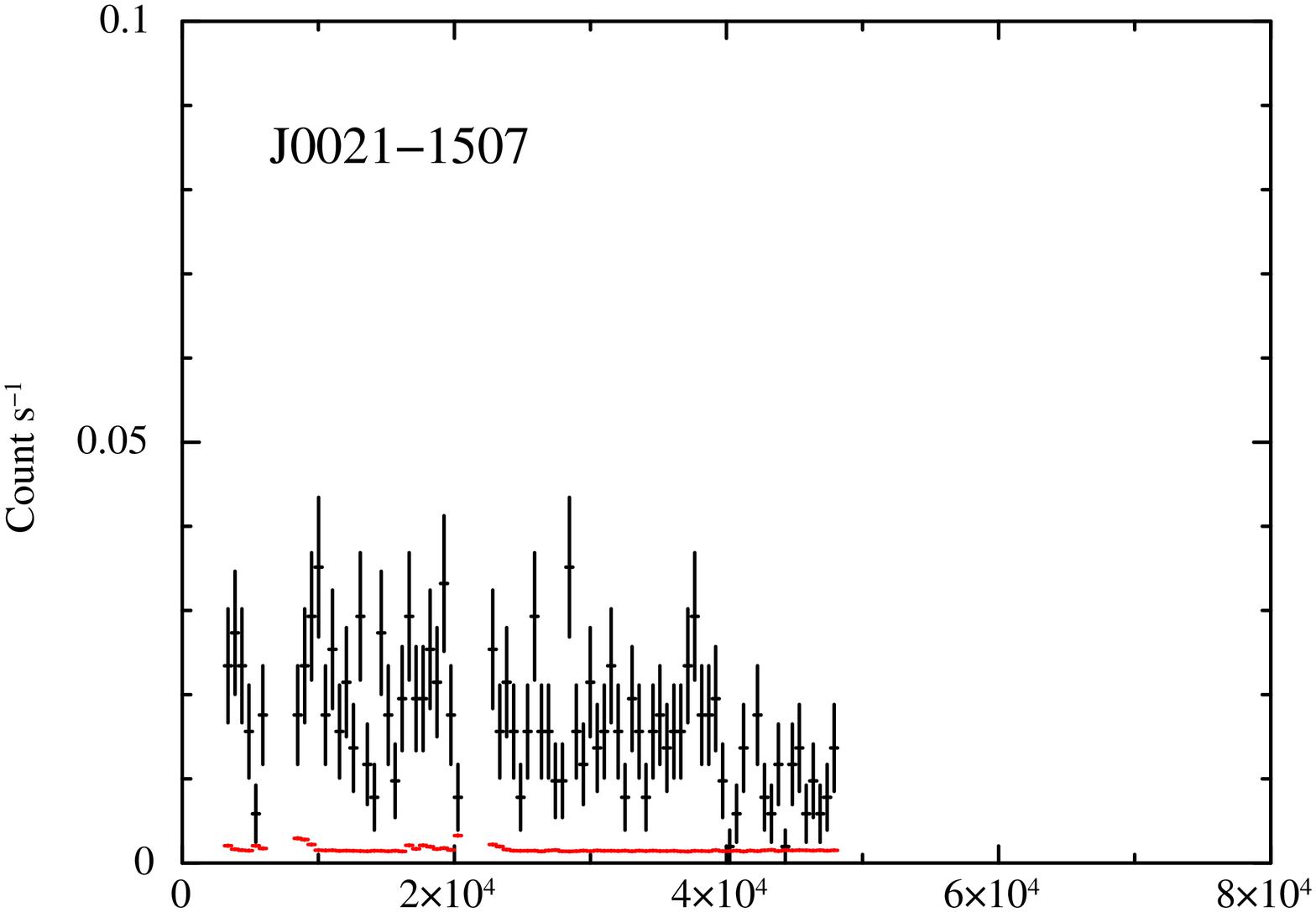}
\includegraphics[width=7.1cm,height=8.25cm,clip,angle=270]{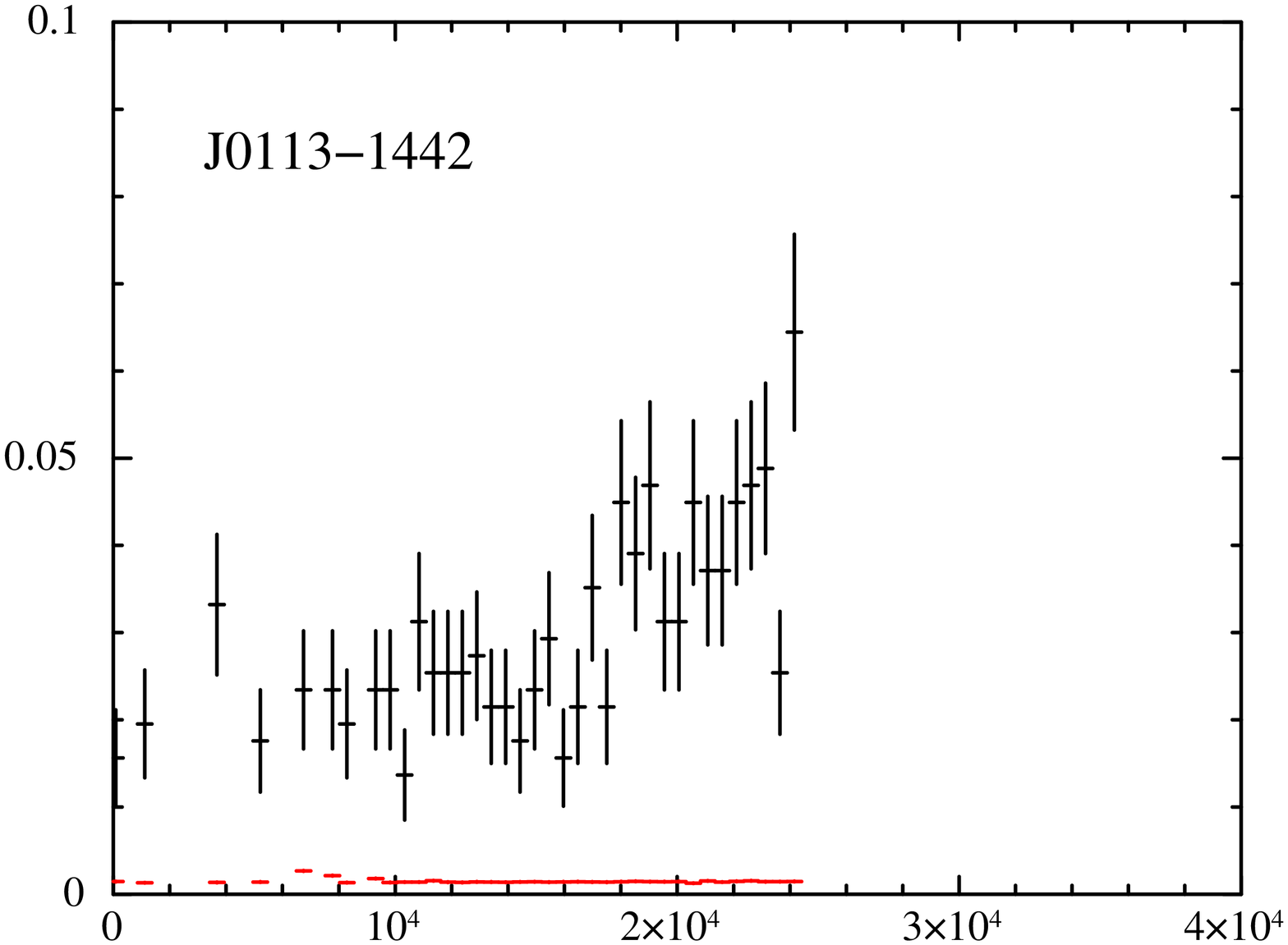}
\includegraphics[width=7.1cm,height=8.75cm,clip,angle=270]{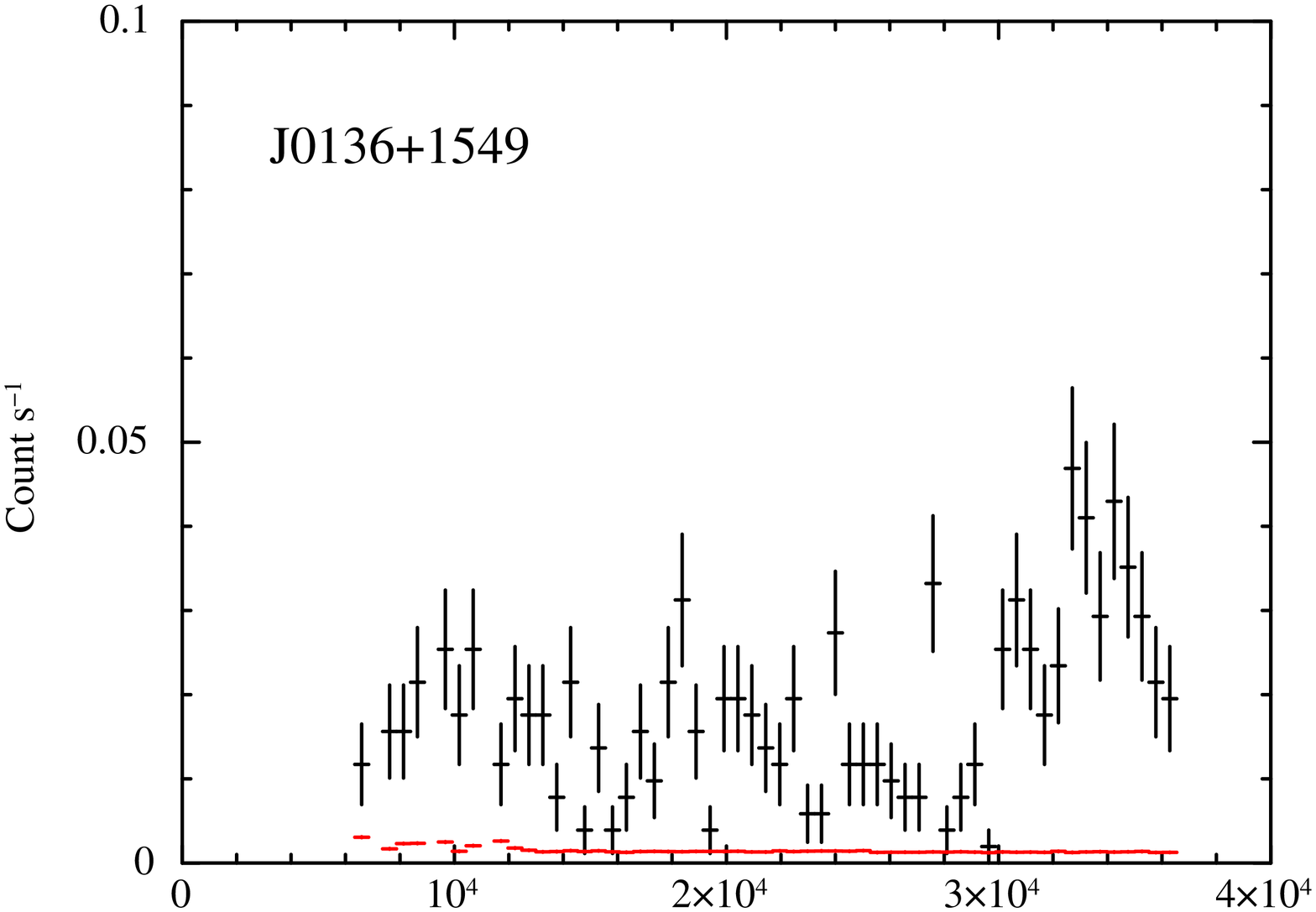}
\includegraphics[width=7.1cm,height=8.25cm,clip,angle=270]{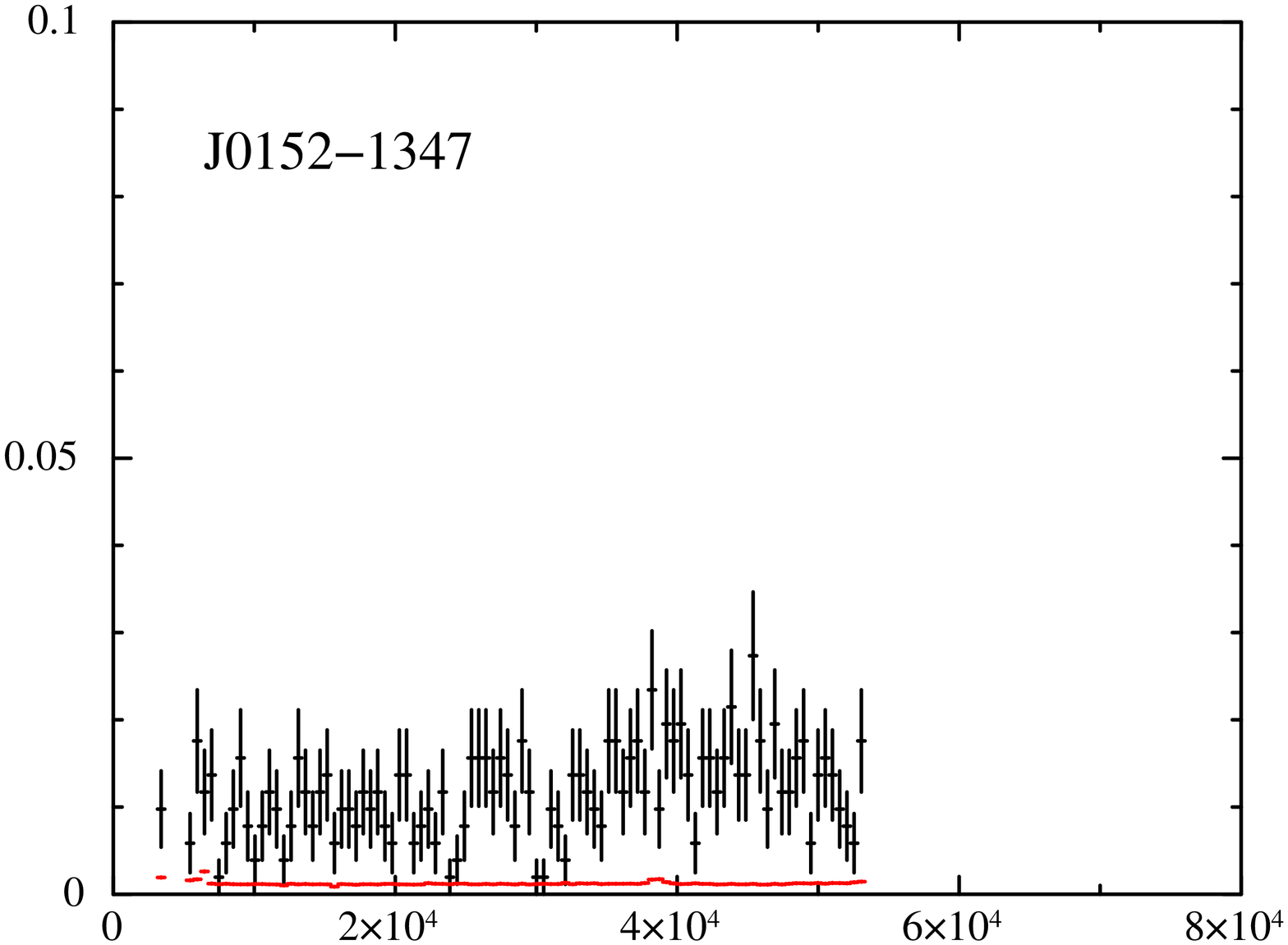}
\includegraphics[width=7.55cm,height=8.75cm,clip,angle=270]{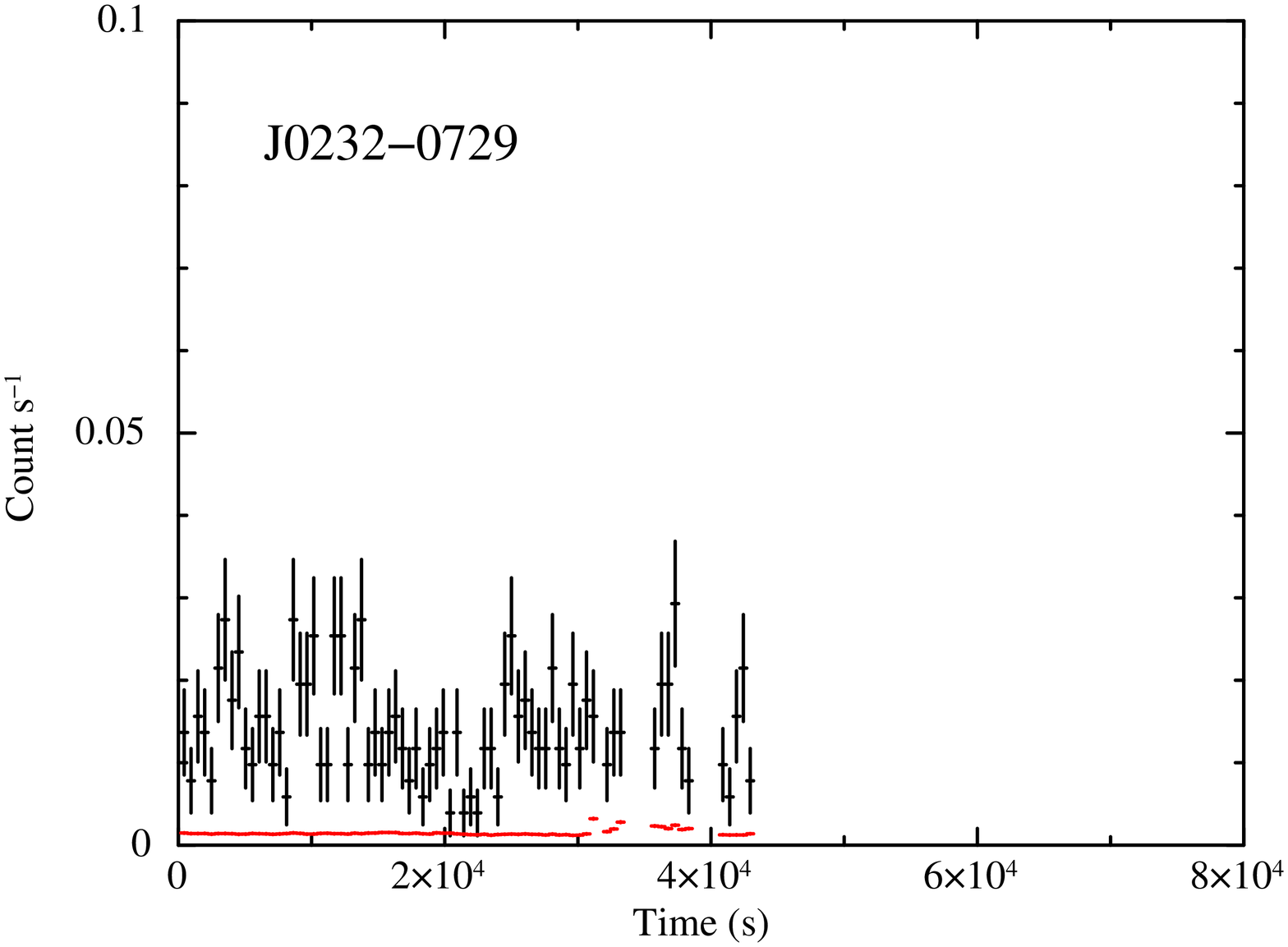}
\hspace{1.15cm}\includegraphics[width=7.55cm,height=8.1cm,clip,angle=270]{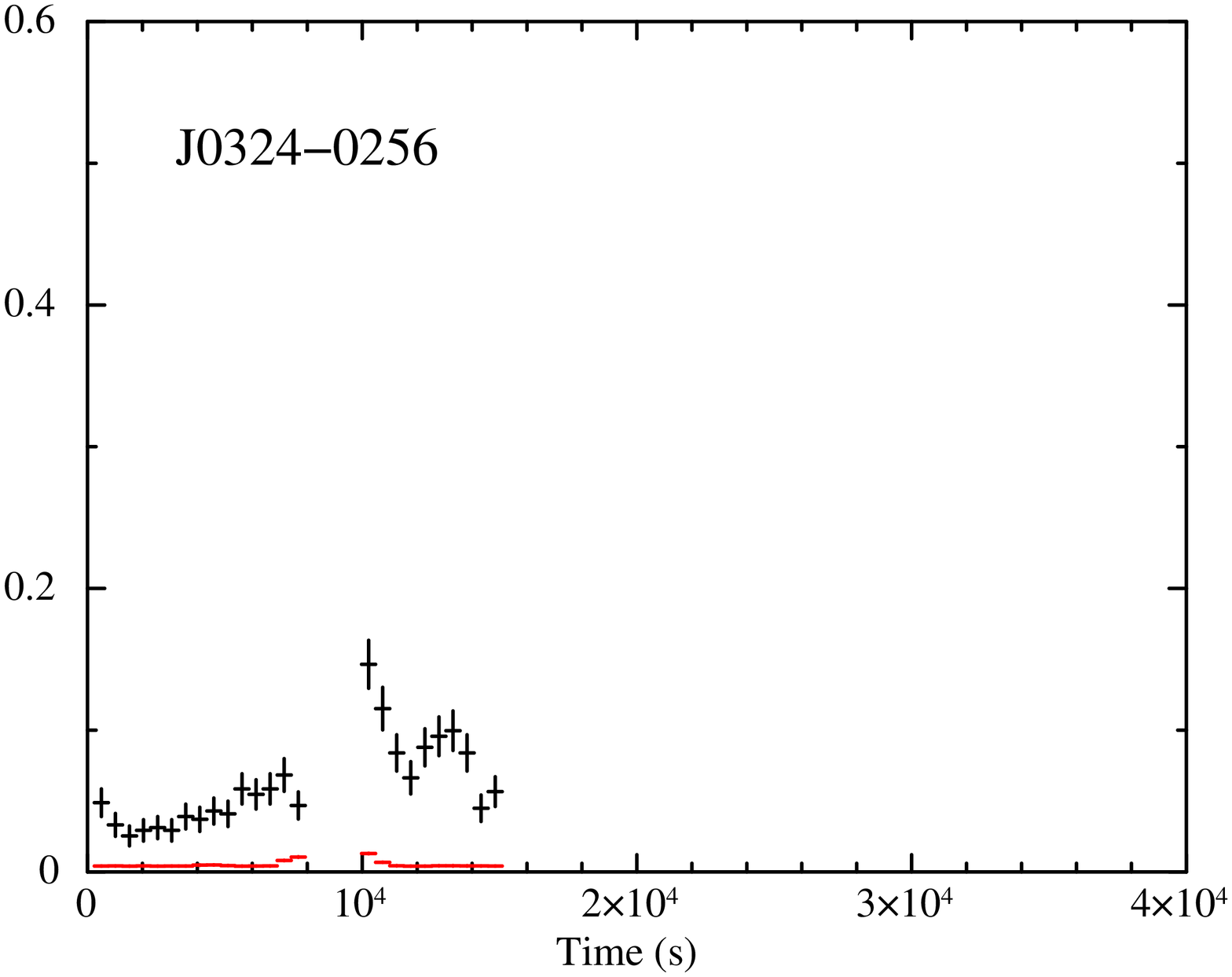}
\caption{Light curves in the 0.5$-$10 keV band with 512 s bins derived from the EPIC-pn data.
Background light curves in the same energy band are over-plotted in each panel (lower data points).}
\label{fig:lc}
\end{figure*}


\begin{figure*}
\figurenum{\ref{fig:lc}}
\includegraphics[width=7.1cm,height=8.75cm,clip,angle=270]{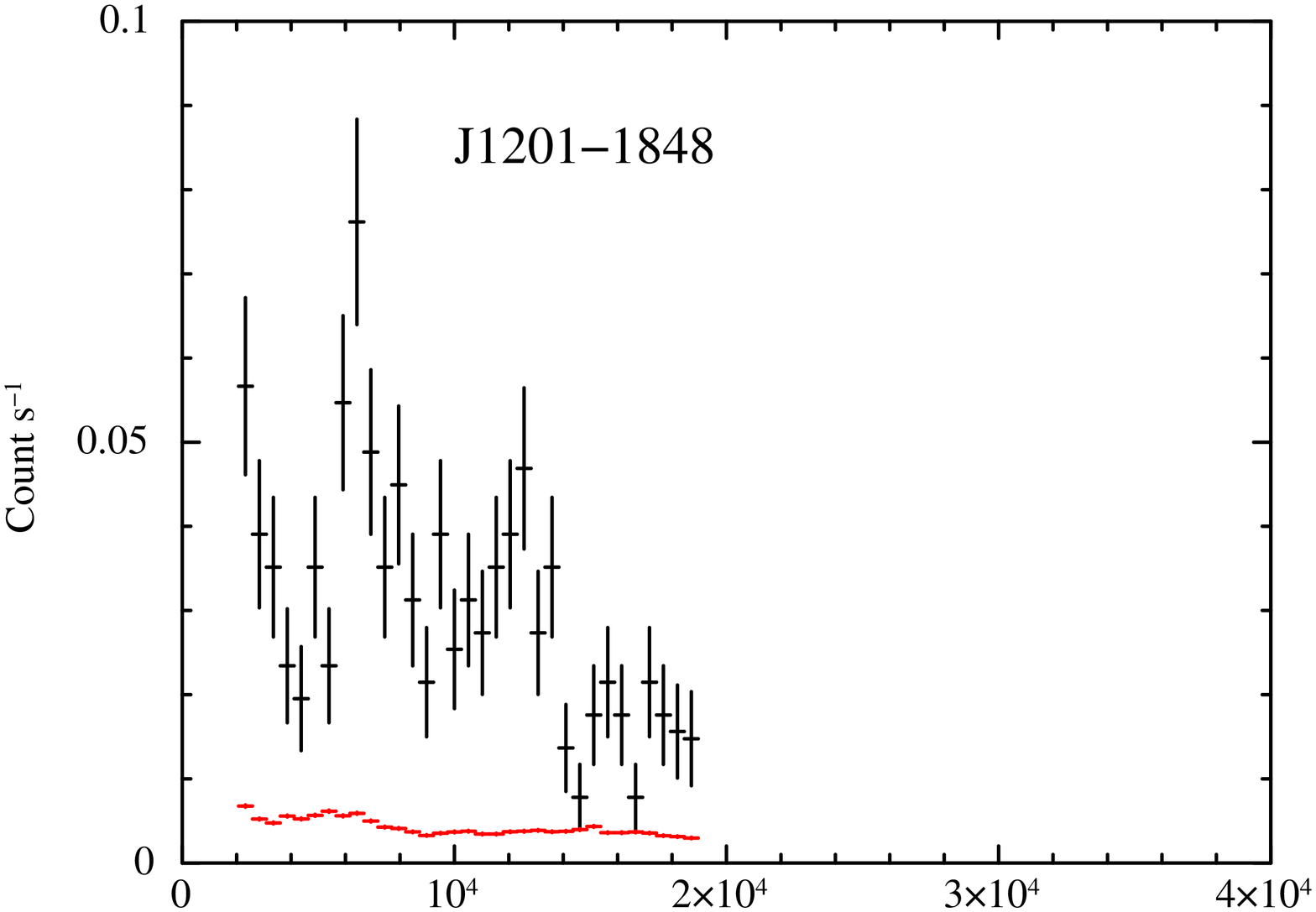}
\includegraphics[width=7.1cm,height=8.25cm,clip,angle=270]{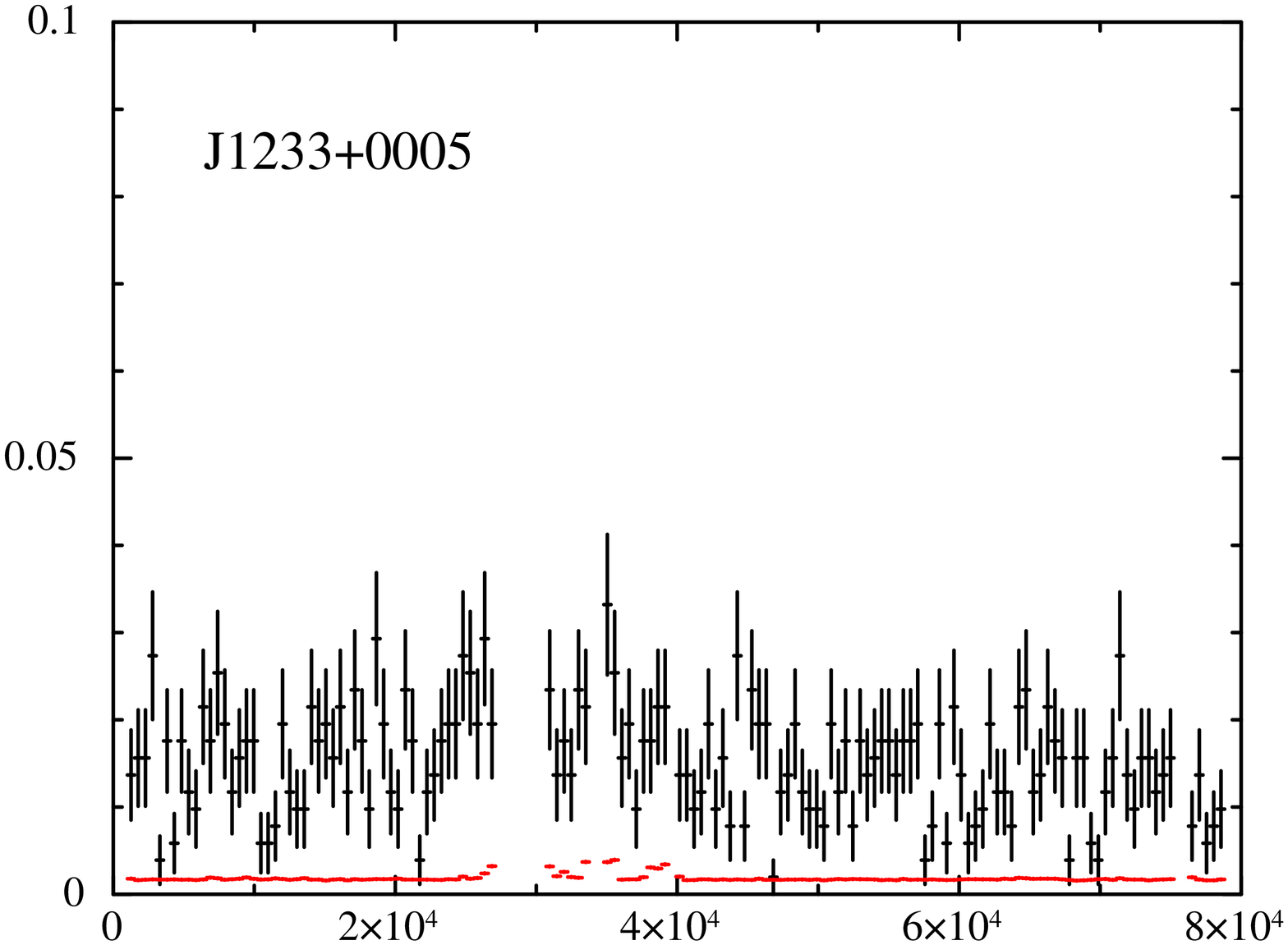}
\includegraphics[width=7.1cm,height=8.75cm,clip,angle=270]{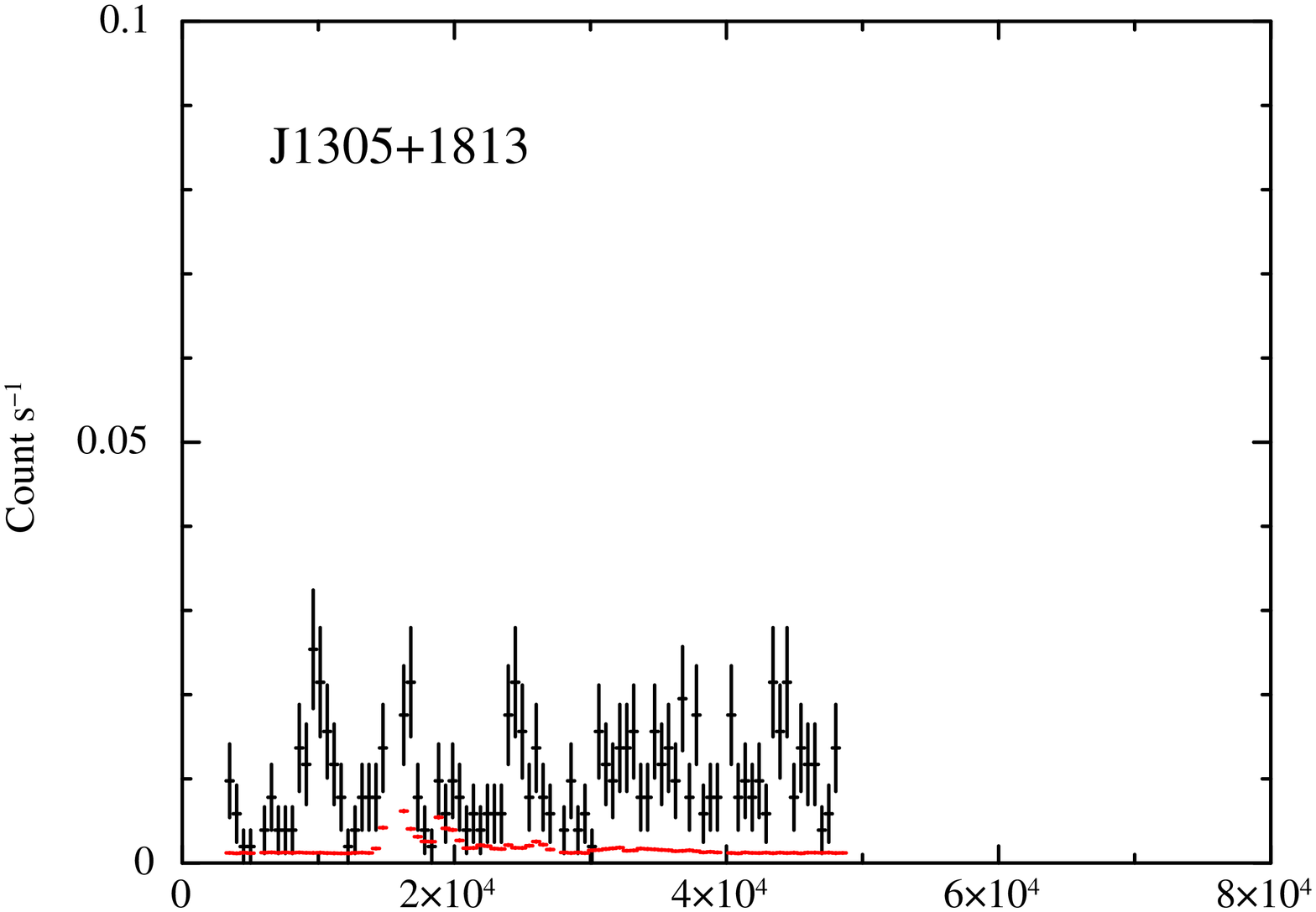}
\includegraphics[width=7.1cm,height=8.25cm,clip,angle=270]{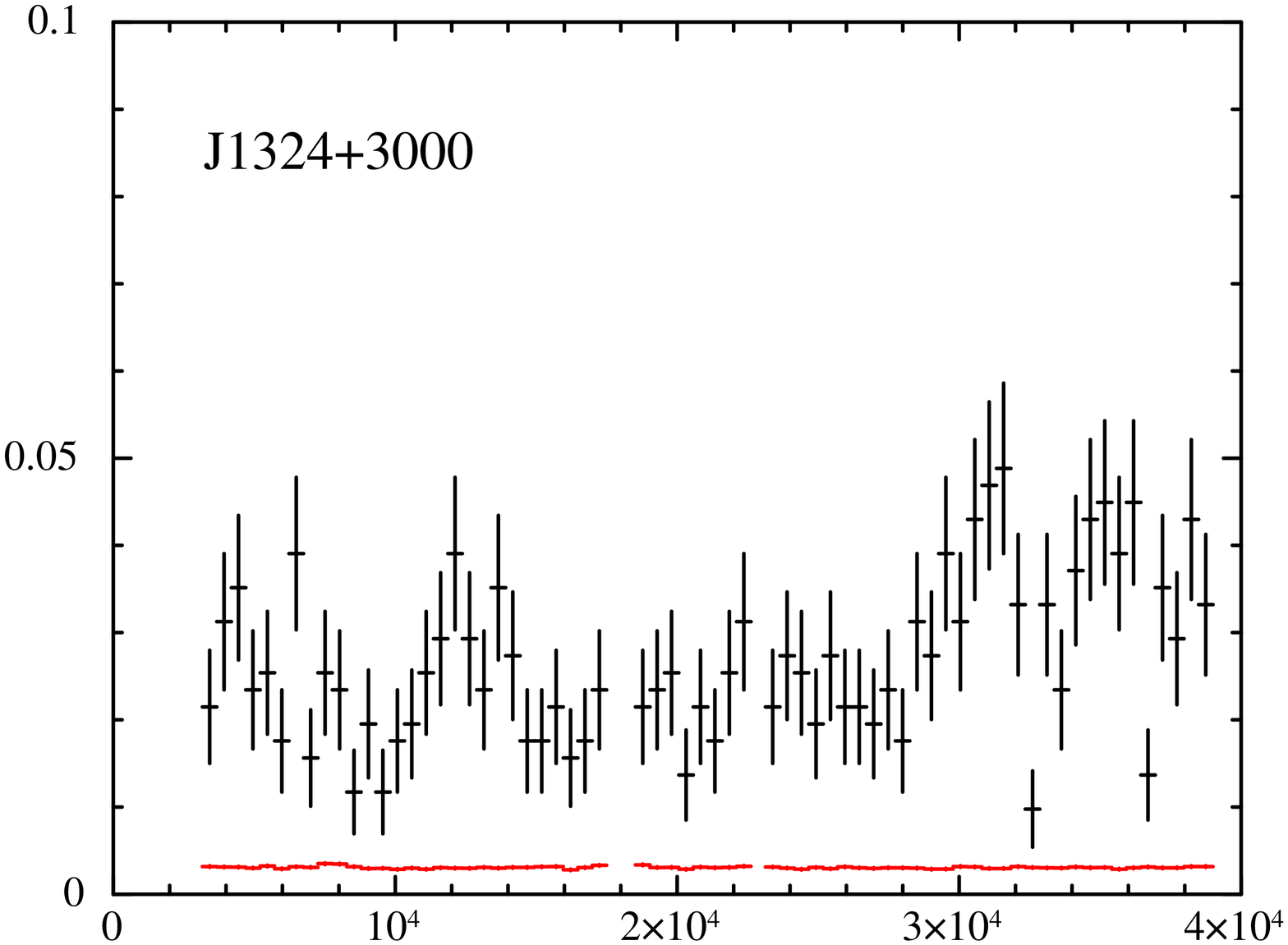}
\includegraphics[width=7.55cm,height=8.75cm,clip,angle=270]{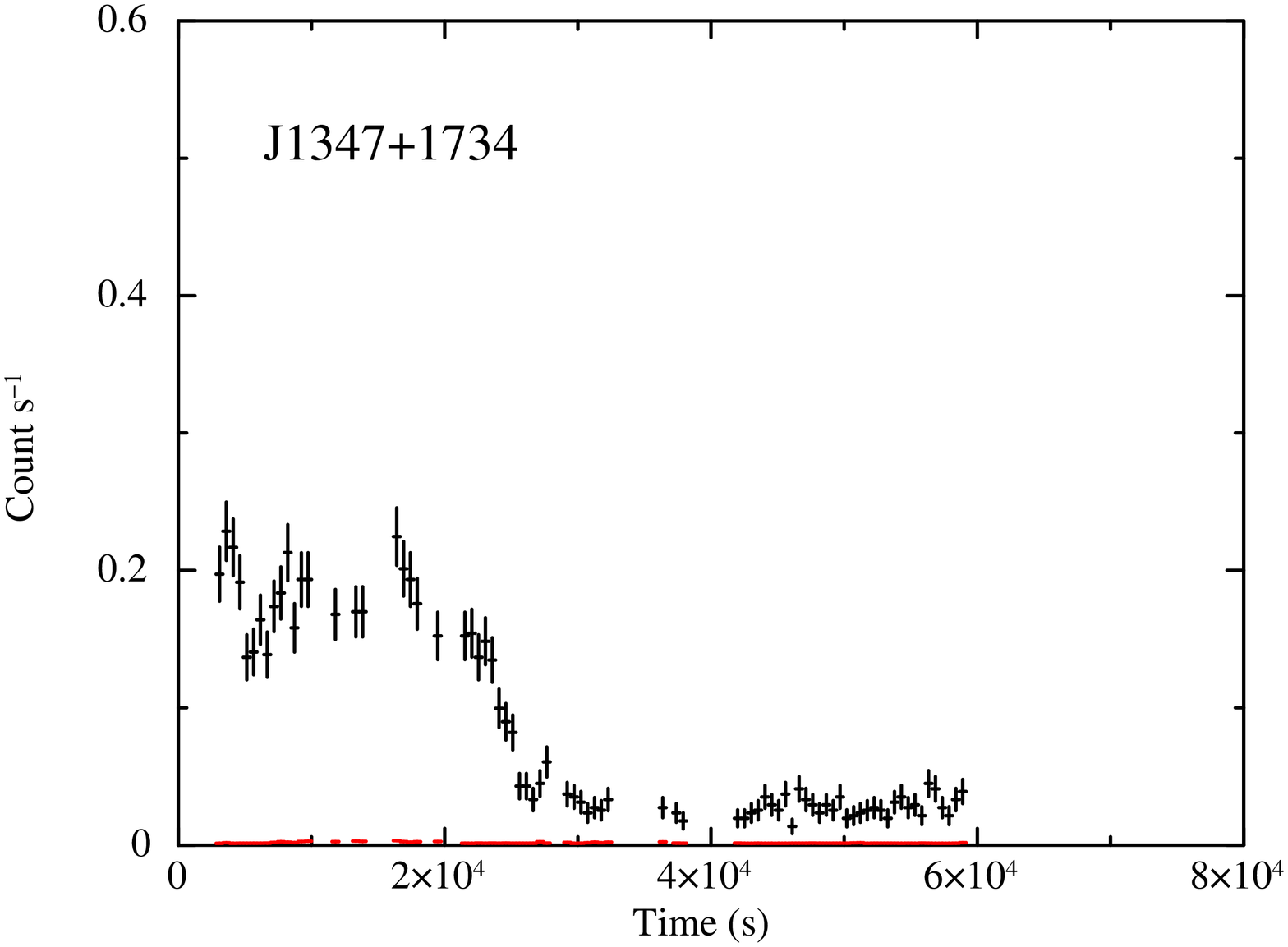}
\hspace{1.15cm}\includegraphics[width=7.55cm,height=8.1cm,clip,angle=270]{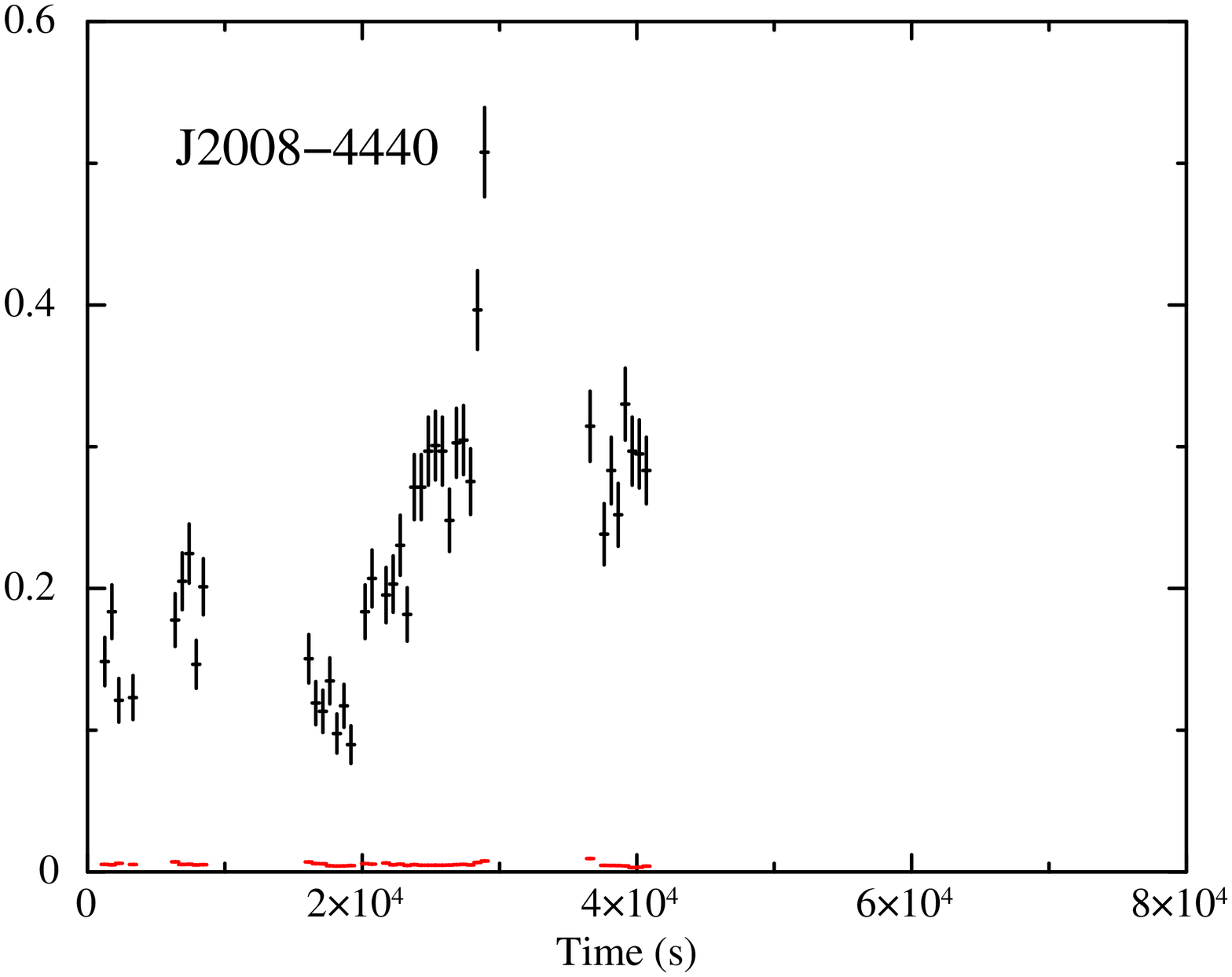}
\caption{Continued}
\end{figure*}


\begin{figure*}
\vspace{1cm}
\figurenum{\ref{fig:lc}}
\includegraphics[width=7.1cm,height=8.75cm,clip,angle=270]{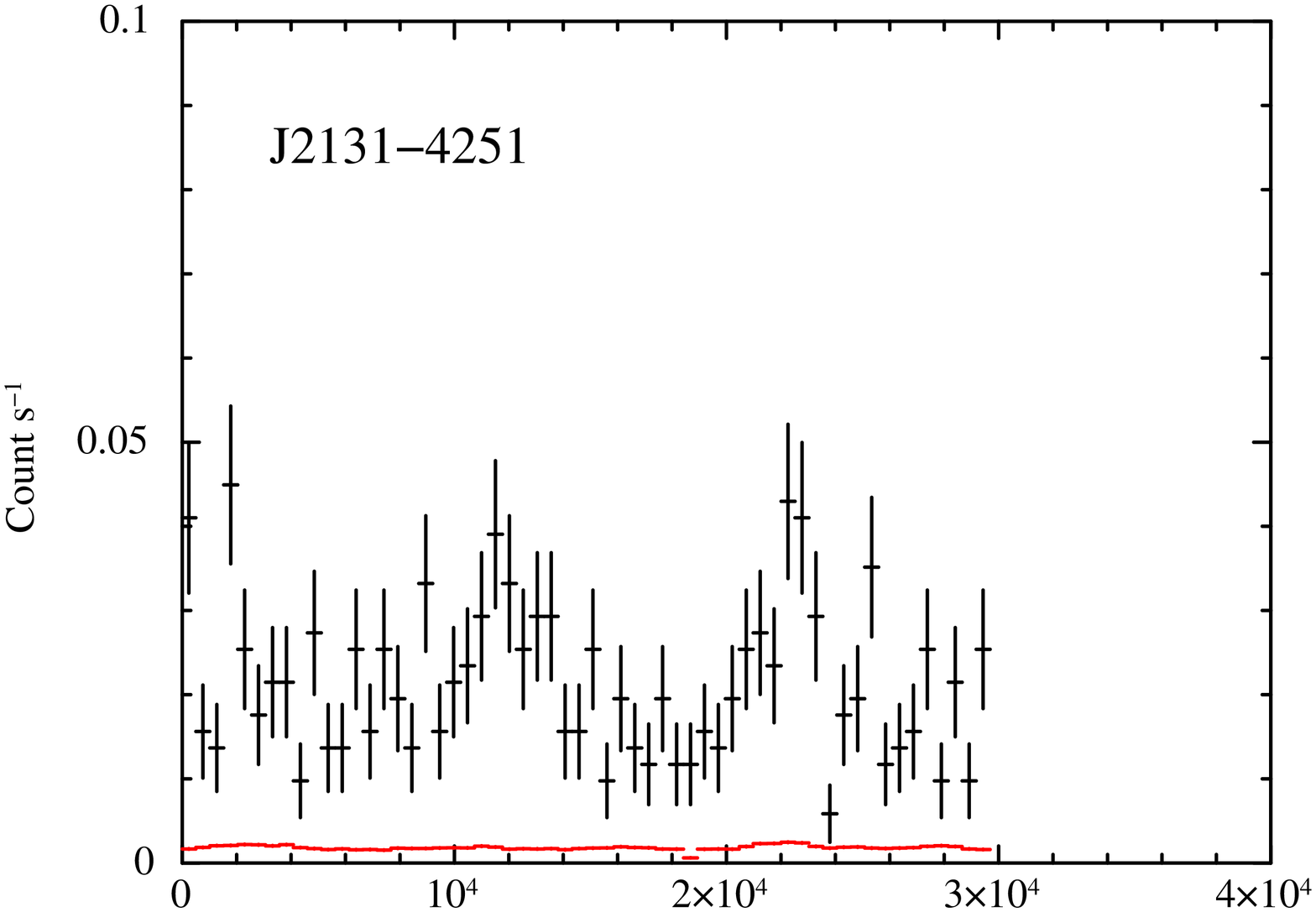}
\includegraphics[width=7.1cm,height=8.1cm,clip,angle=270]{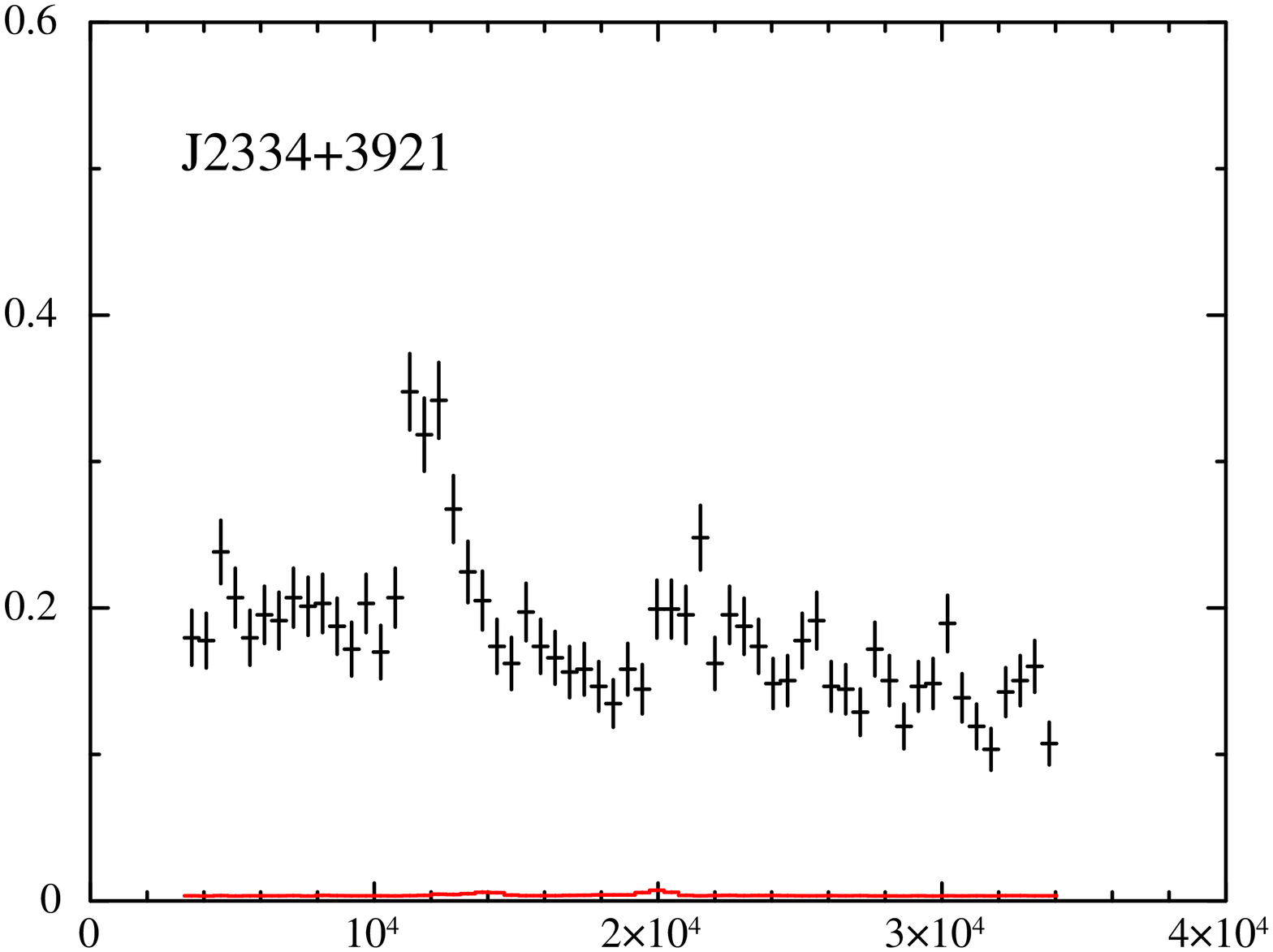}
\includegraphics[width=7.55cm,height=8.75cm,clip,angle=270]{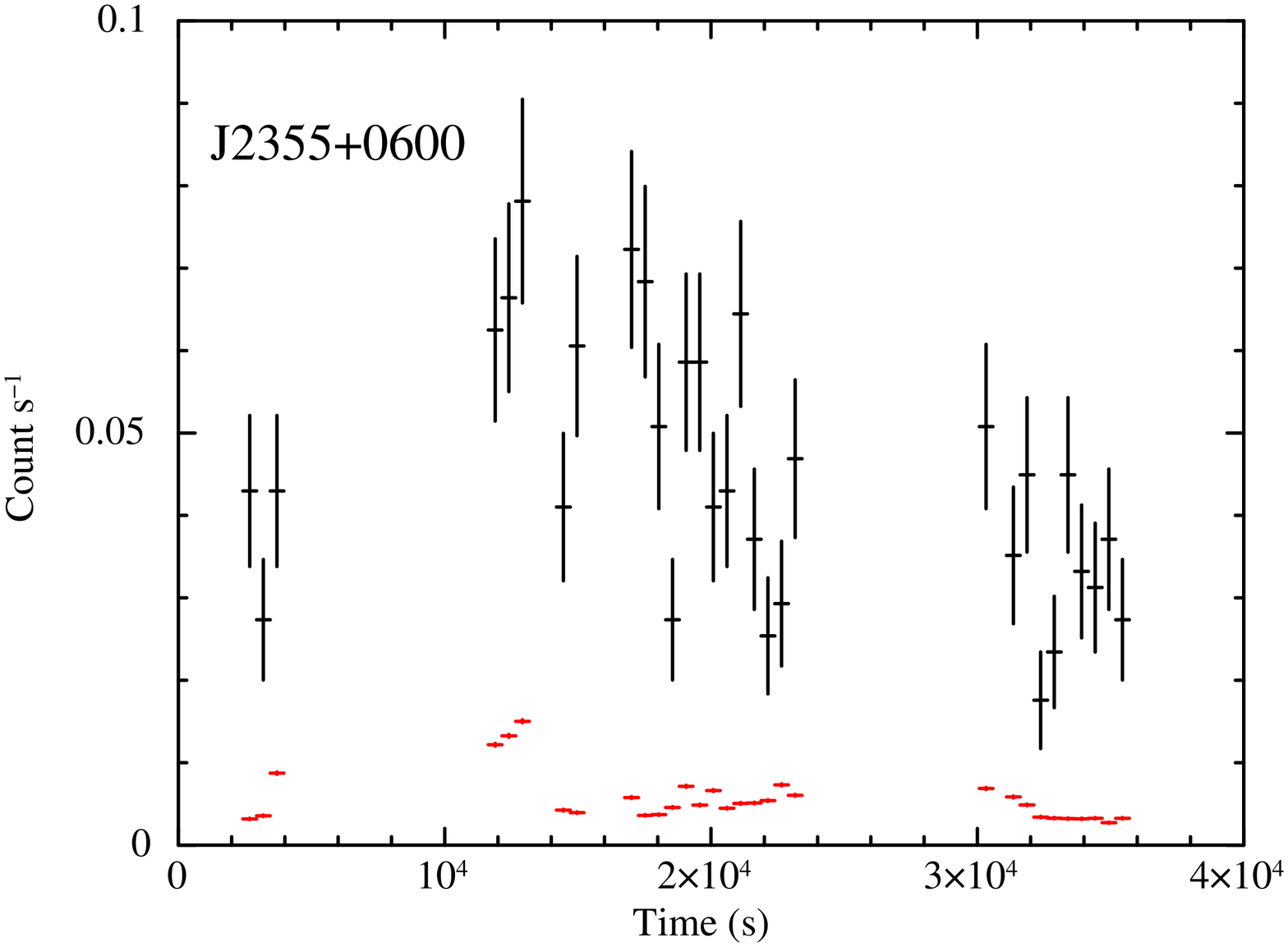}
\caption{Continued}
\end{figure*}




\begin{figure*}
\includegraphics[width=7.1cm,height=8.75cm,clip,angle=270]{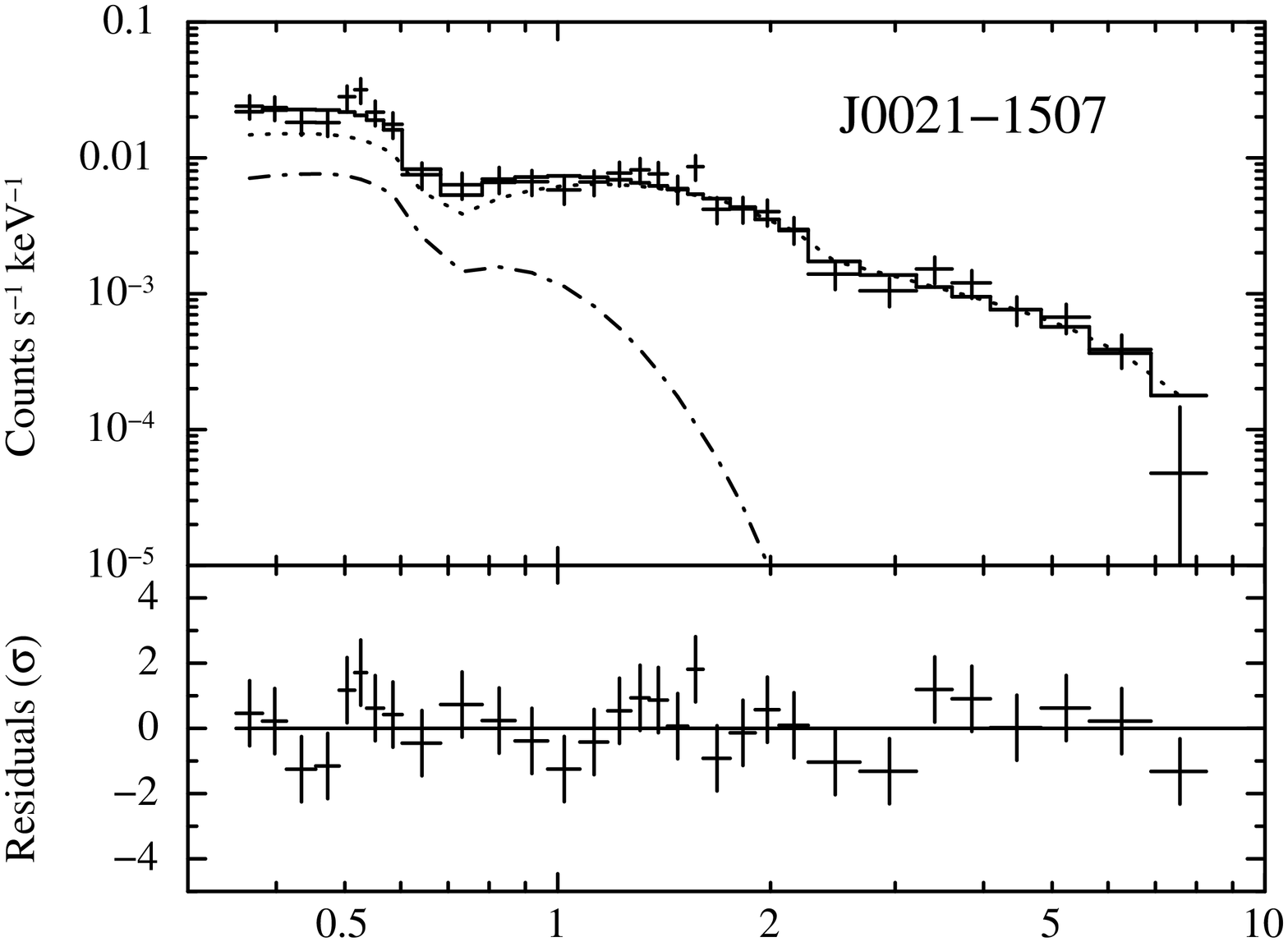}
\includegraphics[width=7.1cm,height=8.25cm,clip,angle=270]{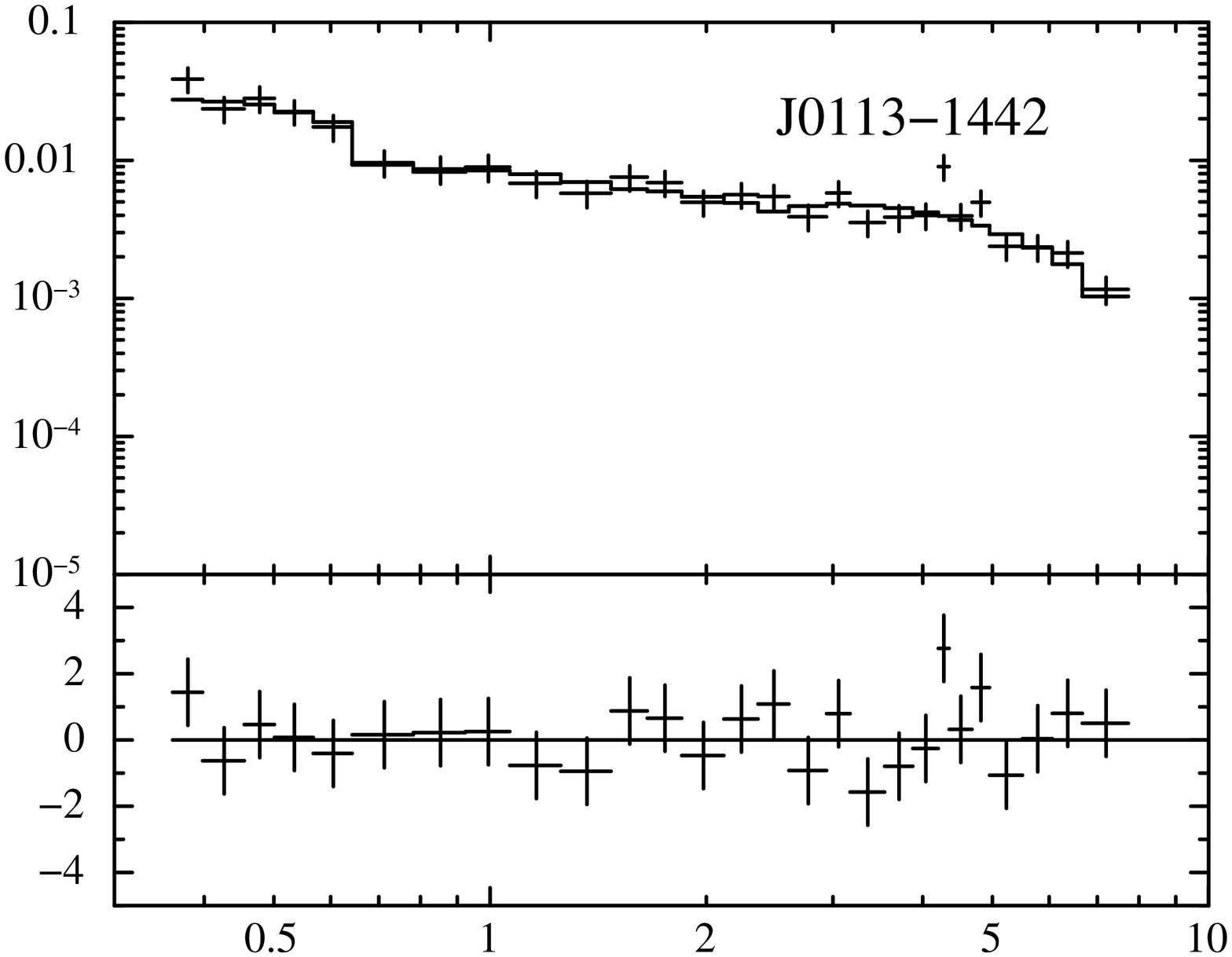}
\includegraphics[width=7.1cm,height=8.75cm,clip,angle=270]{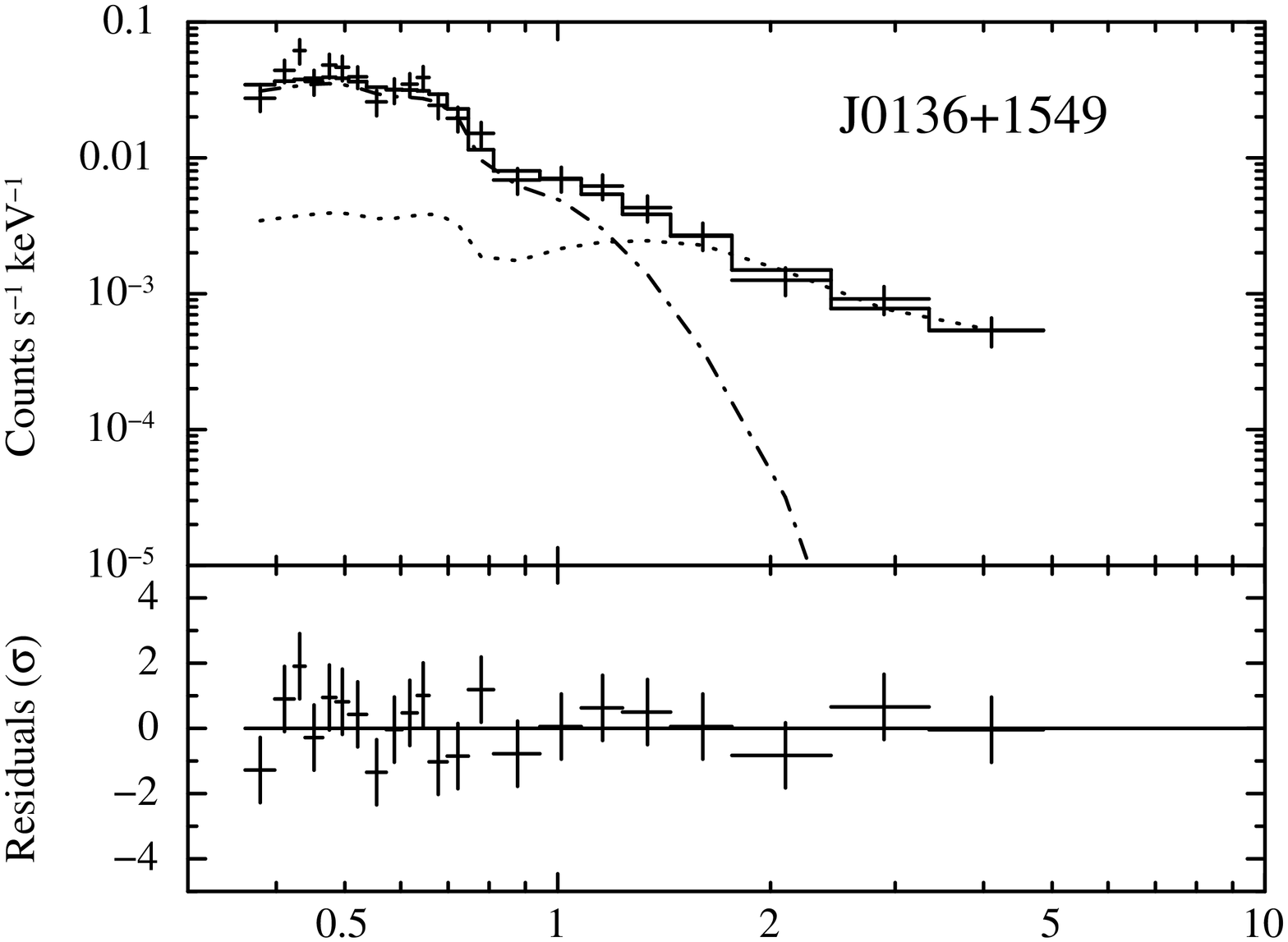}
\includegraphics[width=7.1cm,height=8.25cm,clip,angle=270]{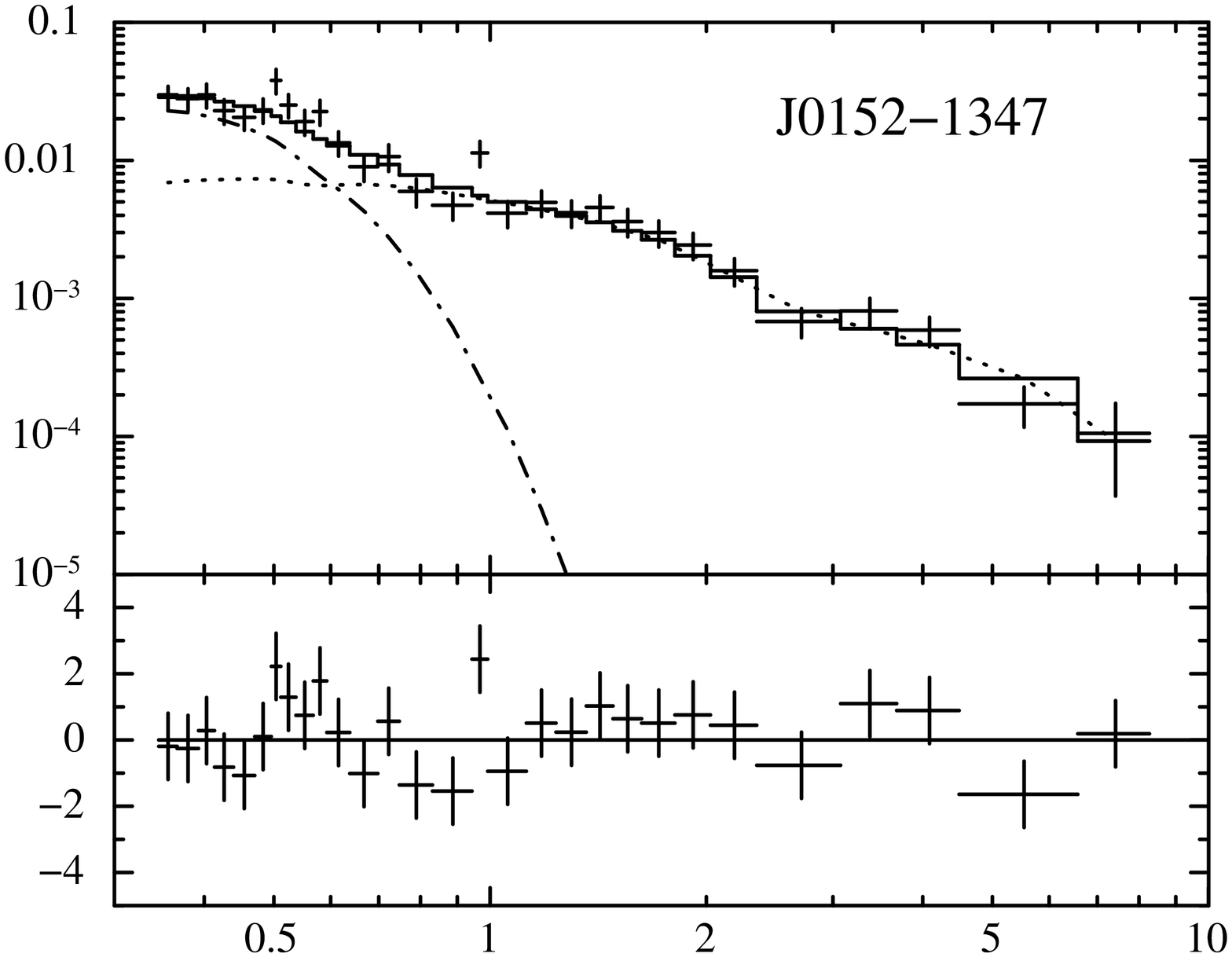}
\includegraphics[width=7.55cm,height=8.75cm,clip,angle=270]{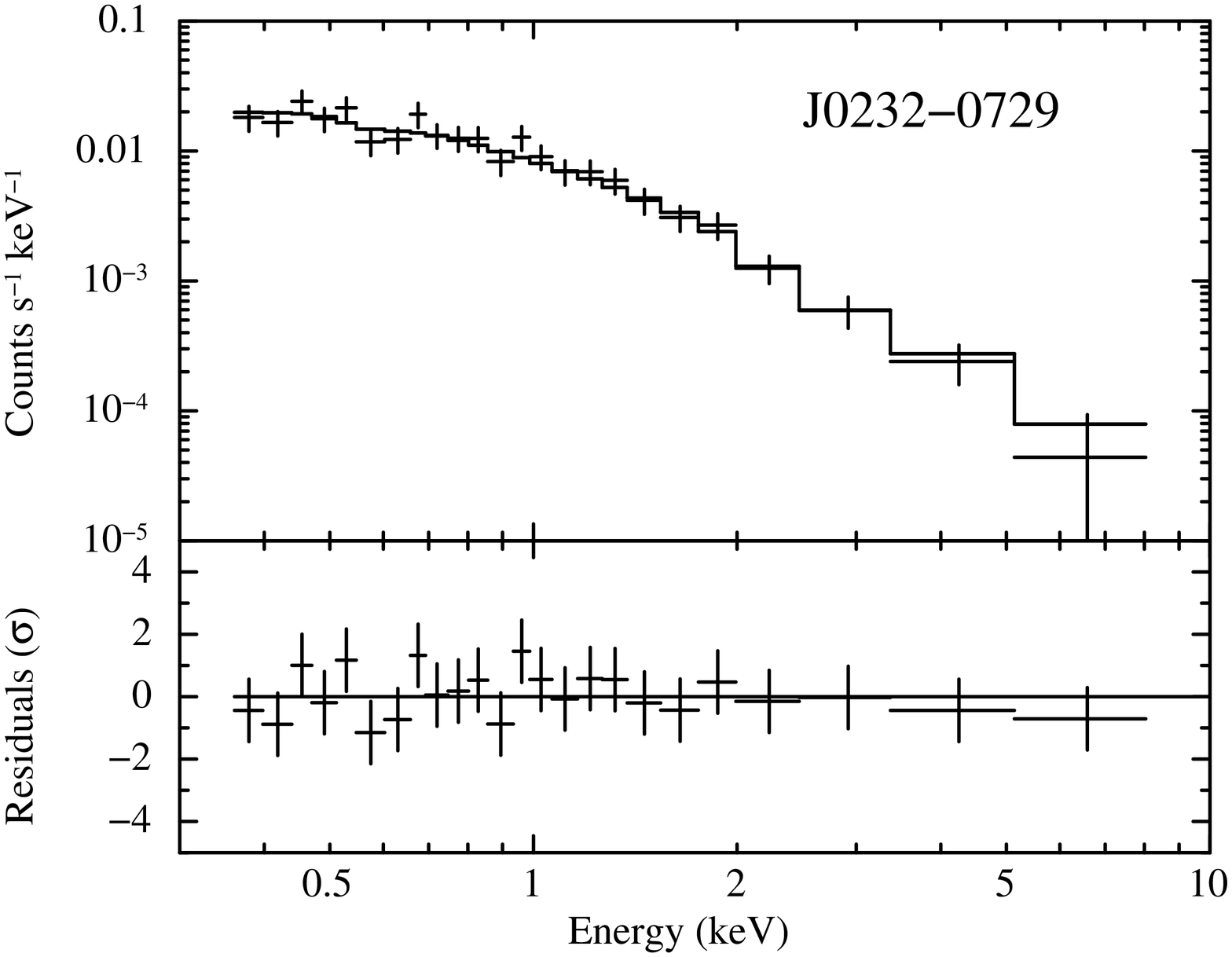}
\hspace{1.0cm}\includegraphics[width=7.55cm,height=8.25cm,clip,angle=270]{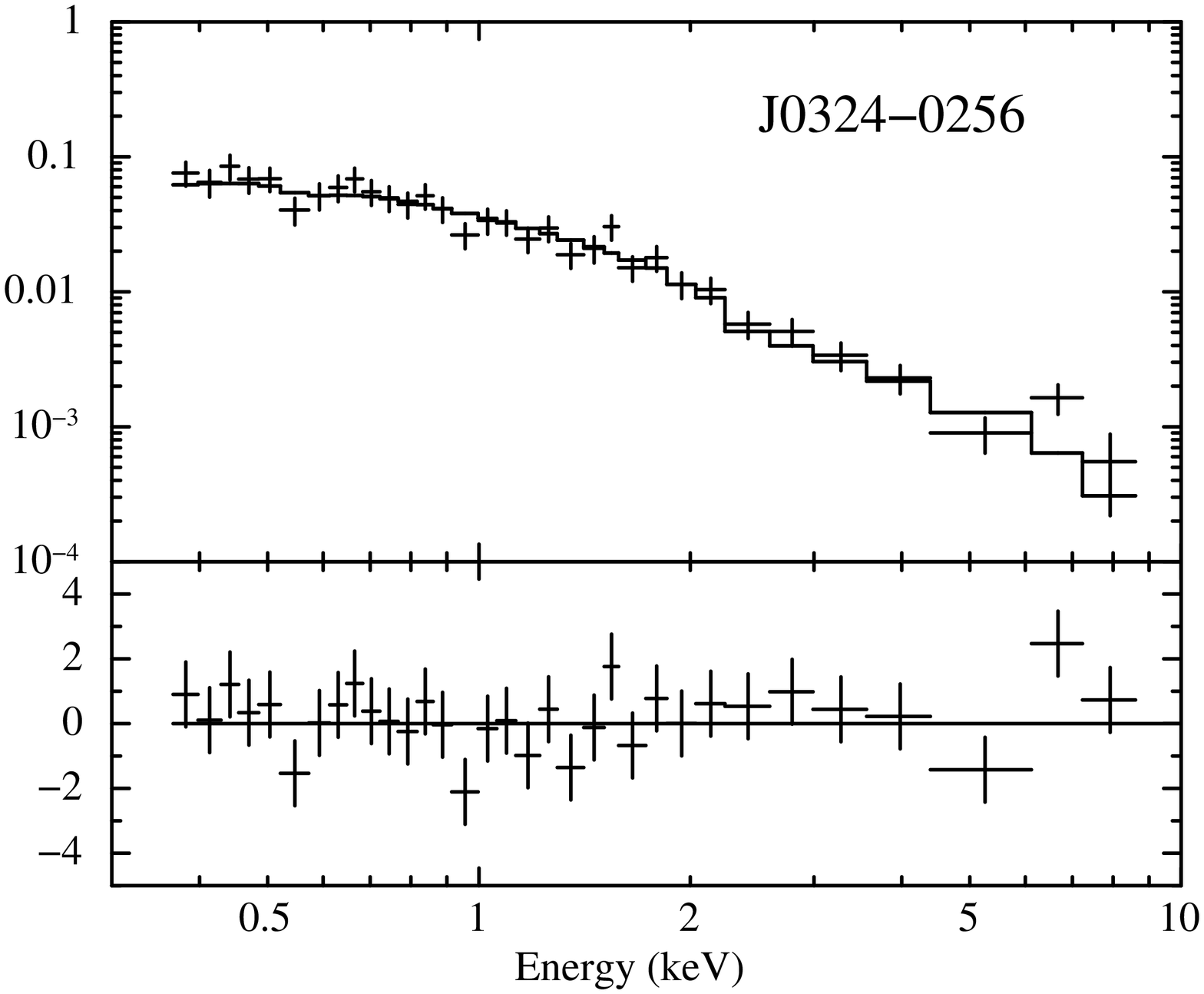}
\caption{X-ray spectra derived from the EPIC-pn data (upper panels) and residuals in units of $\sigma$ (lower panels). Best-fit model is shown as solid histograms. Model components are shown with dashed, dash-dotted, and dotted lines.}
\label{fig:sp}
\end{figure*}

\begin{figure*}
\figurenum{\ref{fig:sp}}
\includegraphics[width=7.1cm,height=8.75cm,clip,angle=270]{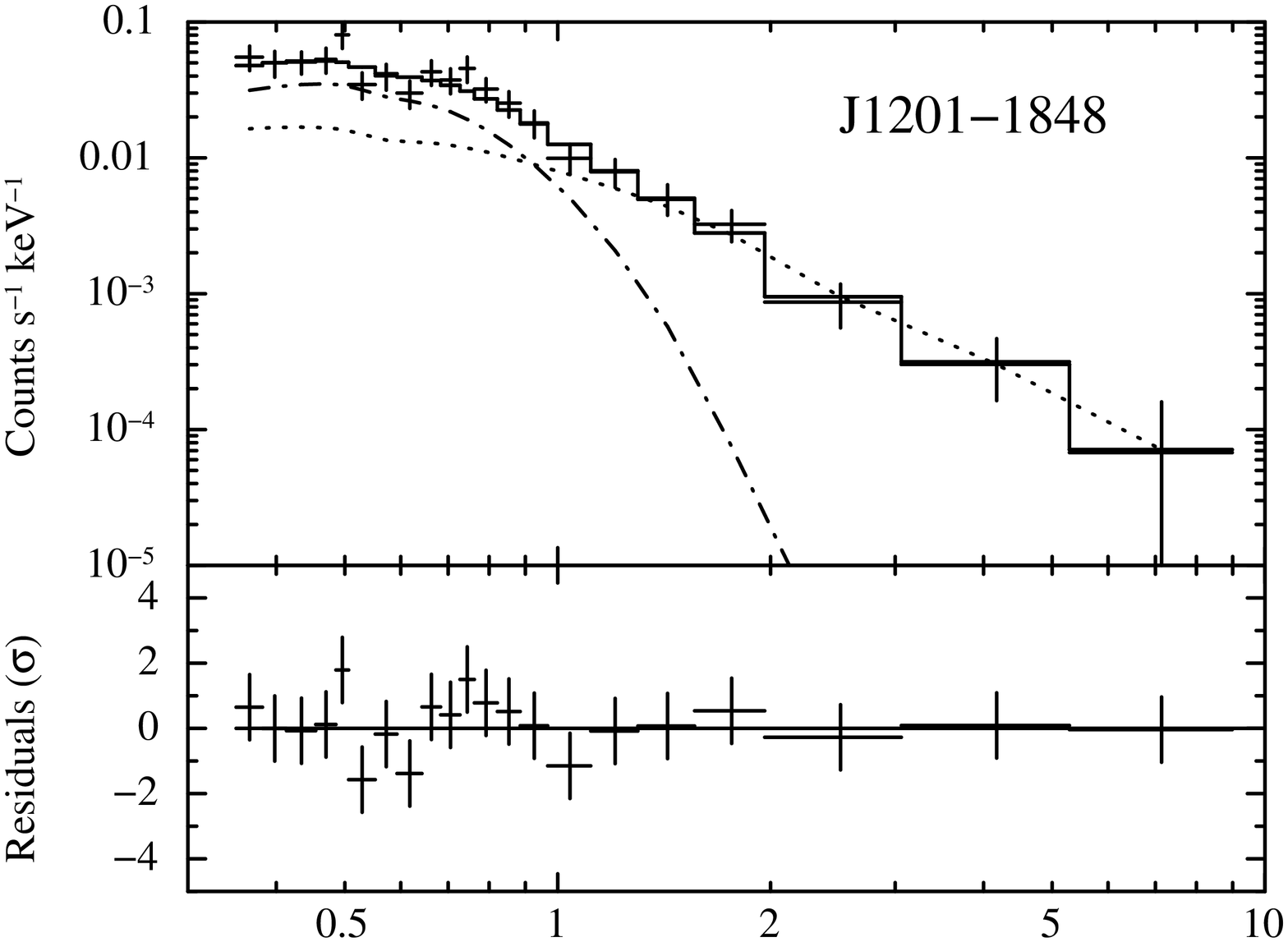}
\includegraphics[width=7.1cm,height=8.25cm,clip,angle=270]{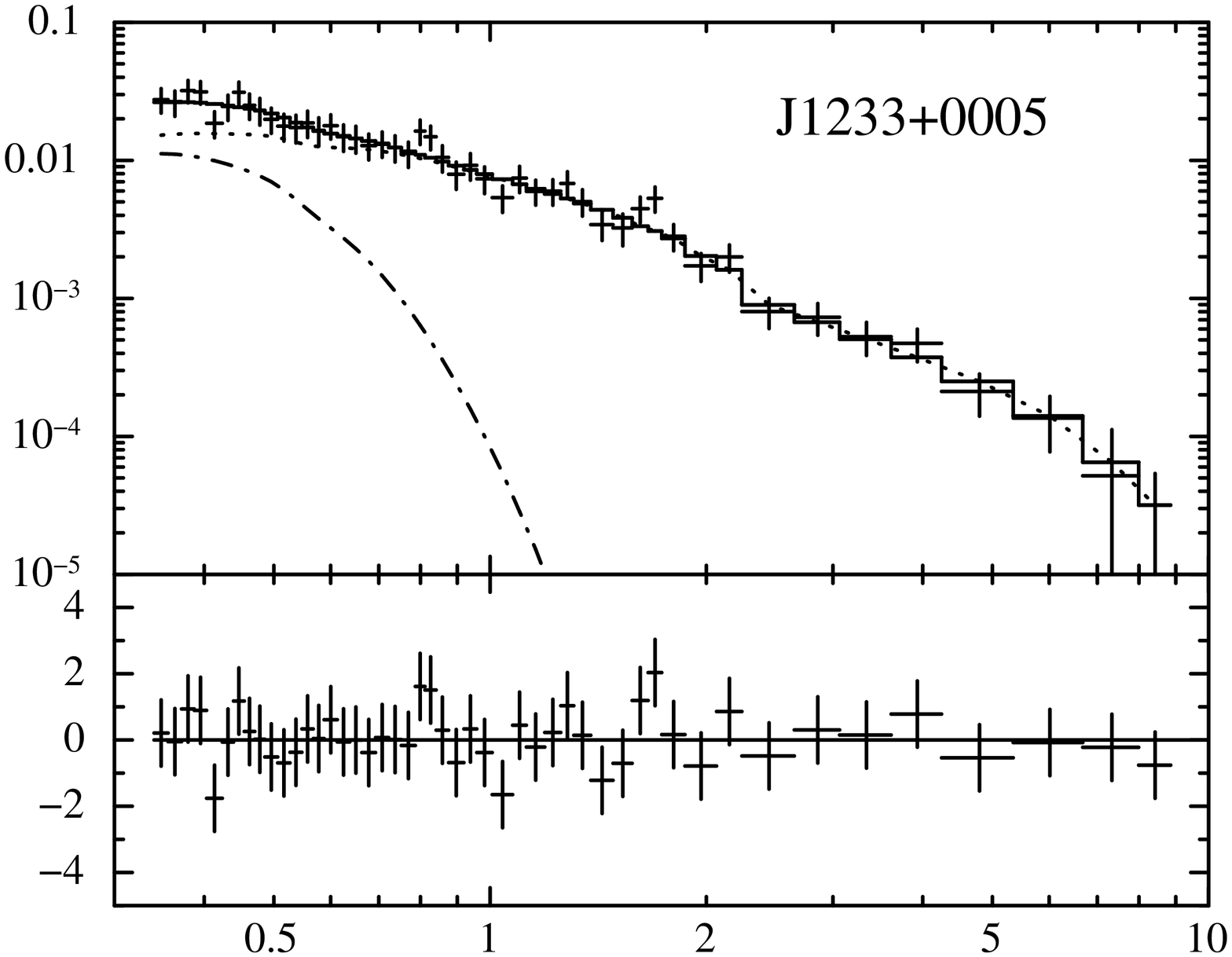}
\includegraphics[width=7.1cm,height=8.75cm,clip,angle=270]{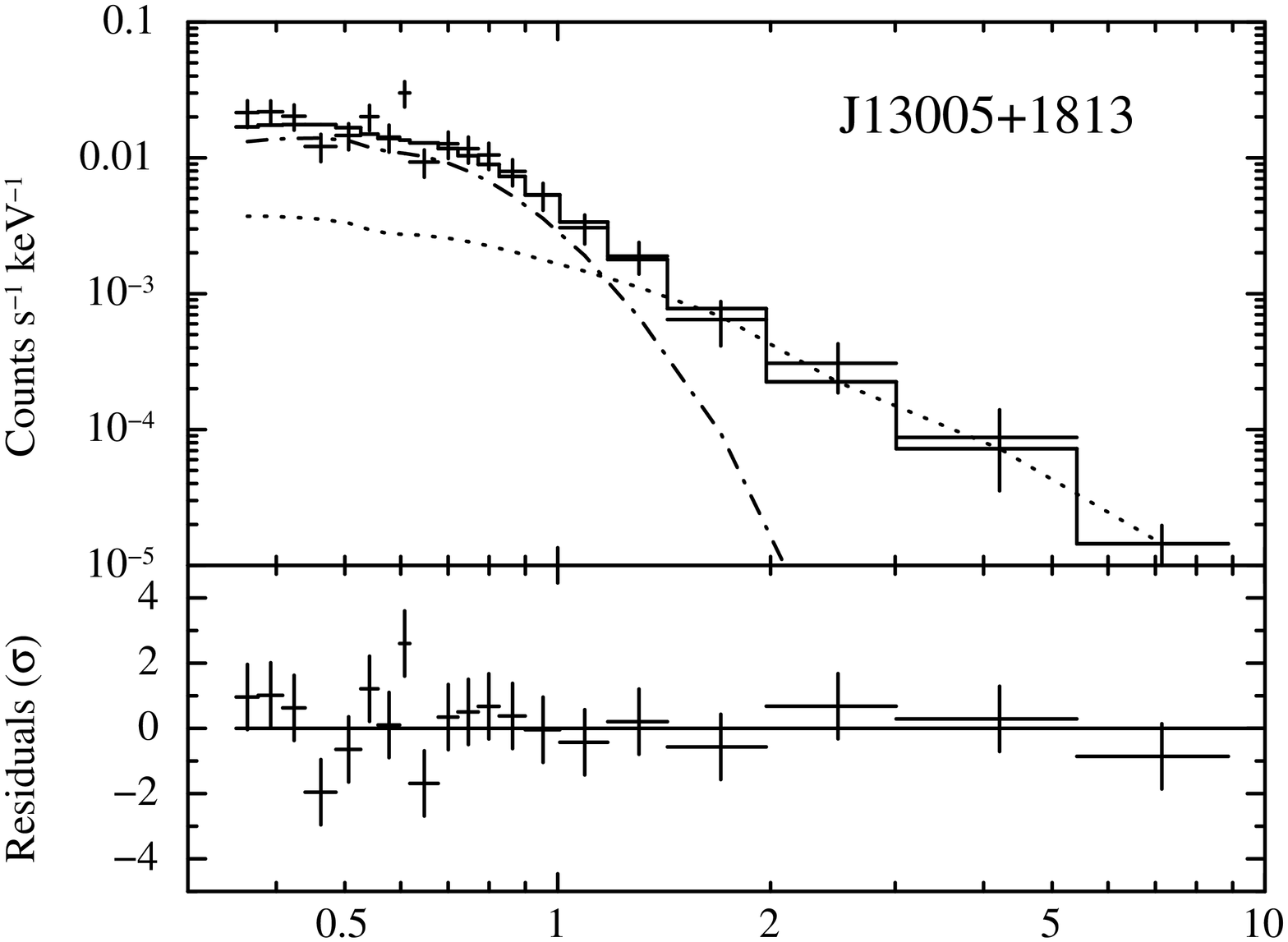}
\includegraphics[width=7.1cm,height=8.25cm,clip,angle=270]{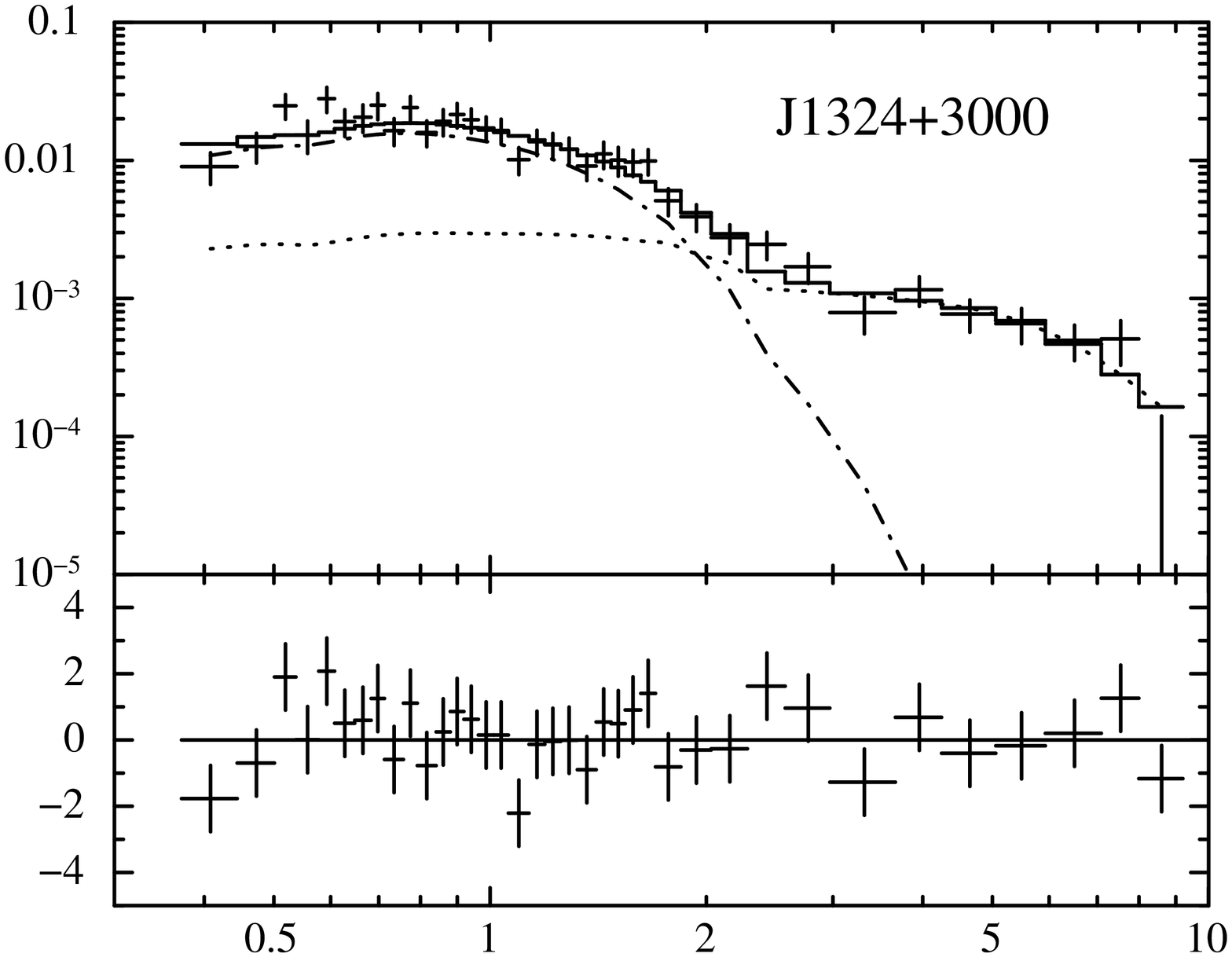}
\includegraphics[width=7.55cm,height=8.75cm,clip,angle=270]{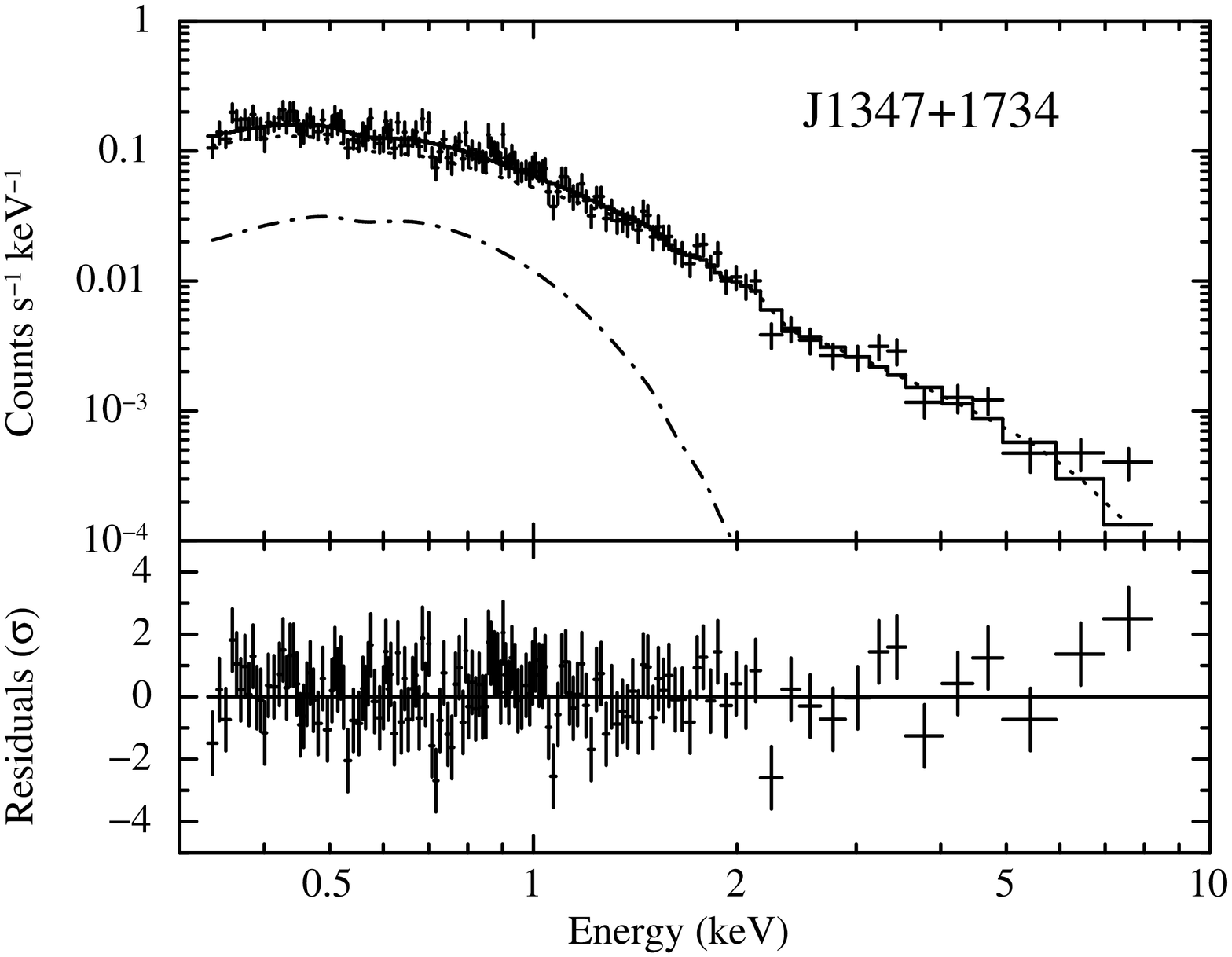}
\hspace{1.0cm}\includegraphics[width=7.55cm,height=8.25cm,clip,angle=270]{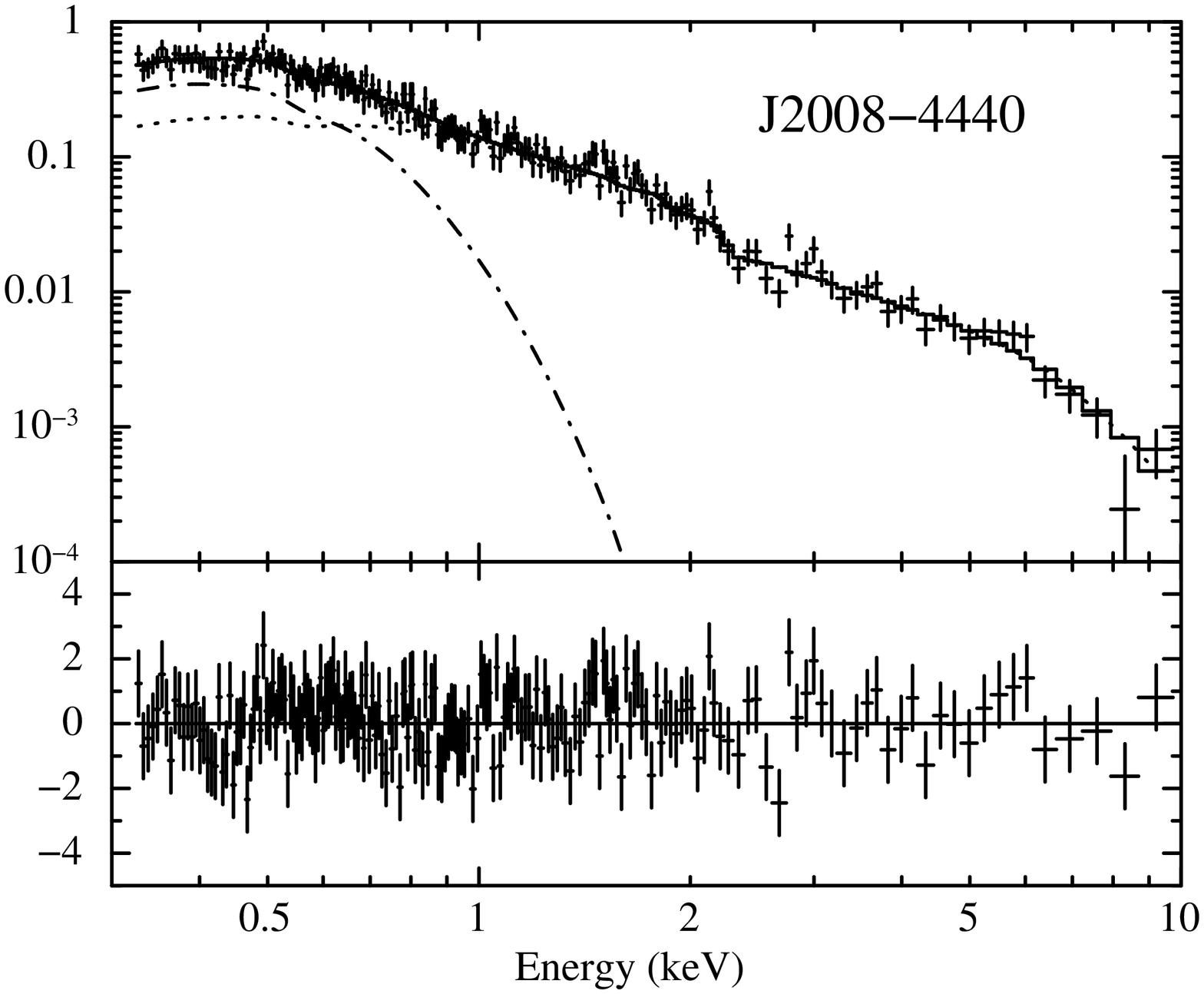}
\caption{Continued}
\end{figure*}

\begin{figure*}
\figurenum{\ref{fig:sp}}
\includegraphics[width=7.1cm,height=8.75cm,clip,angle=270]{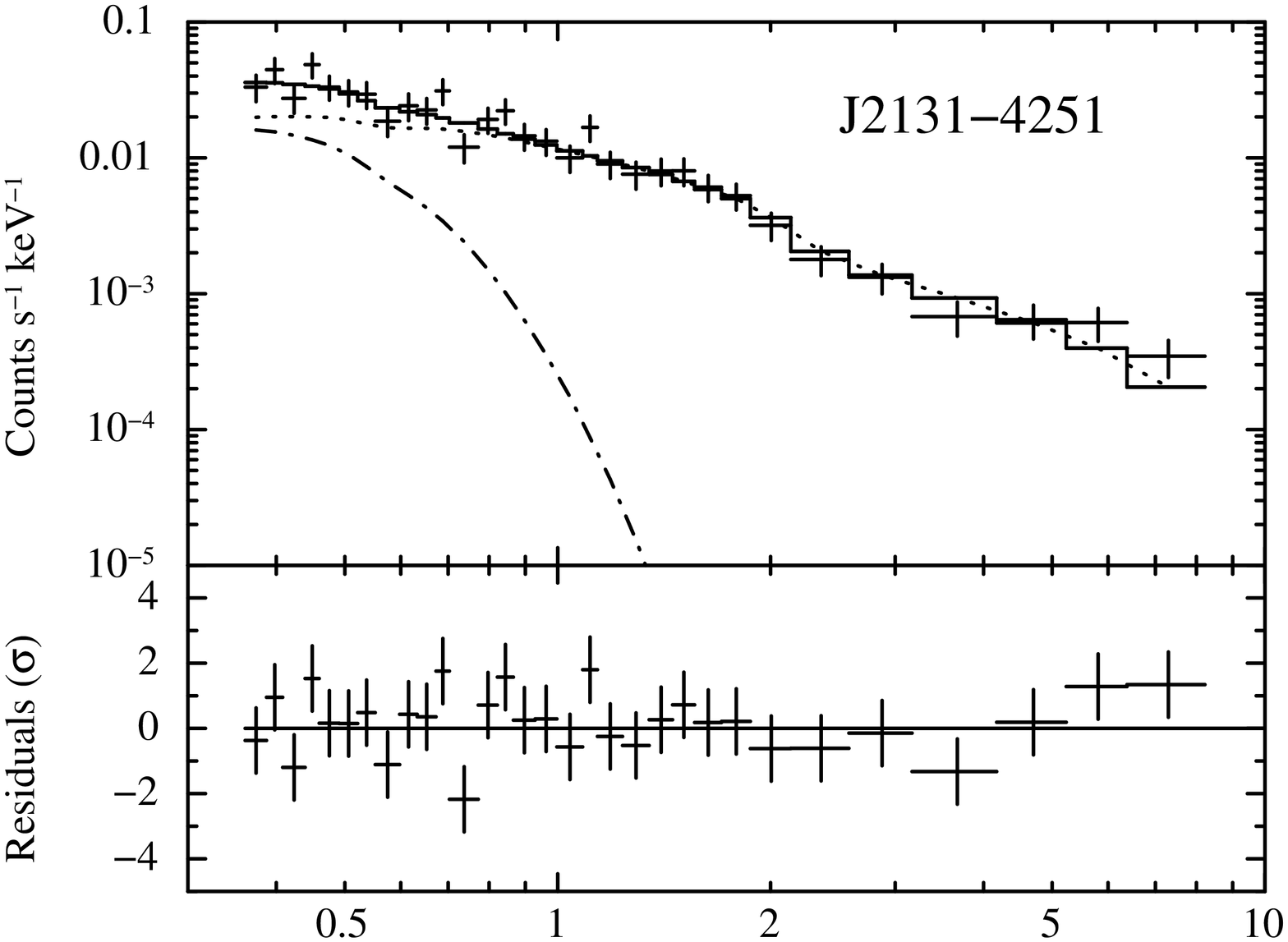}
\includegraphics[width=7.1cm,height=8.25cm,clip,angle=270]{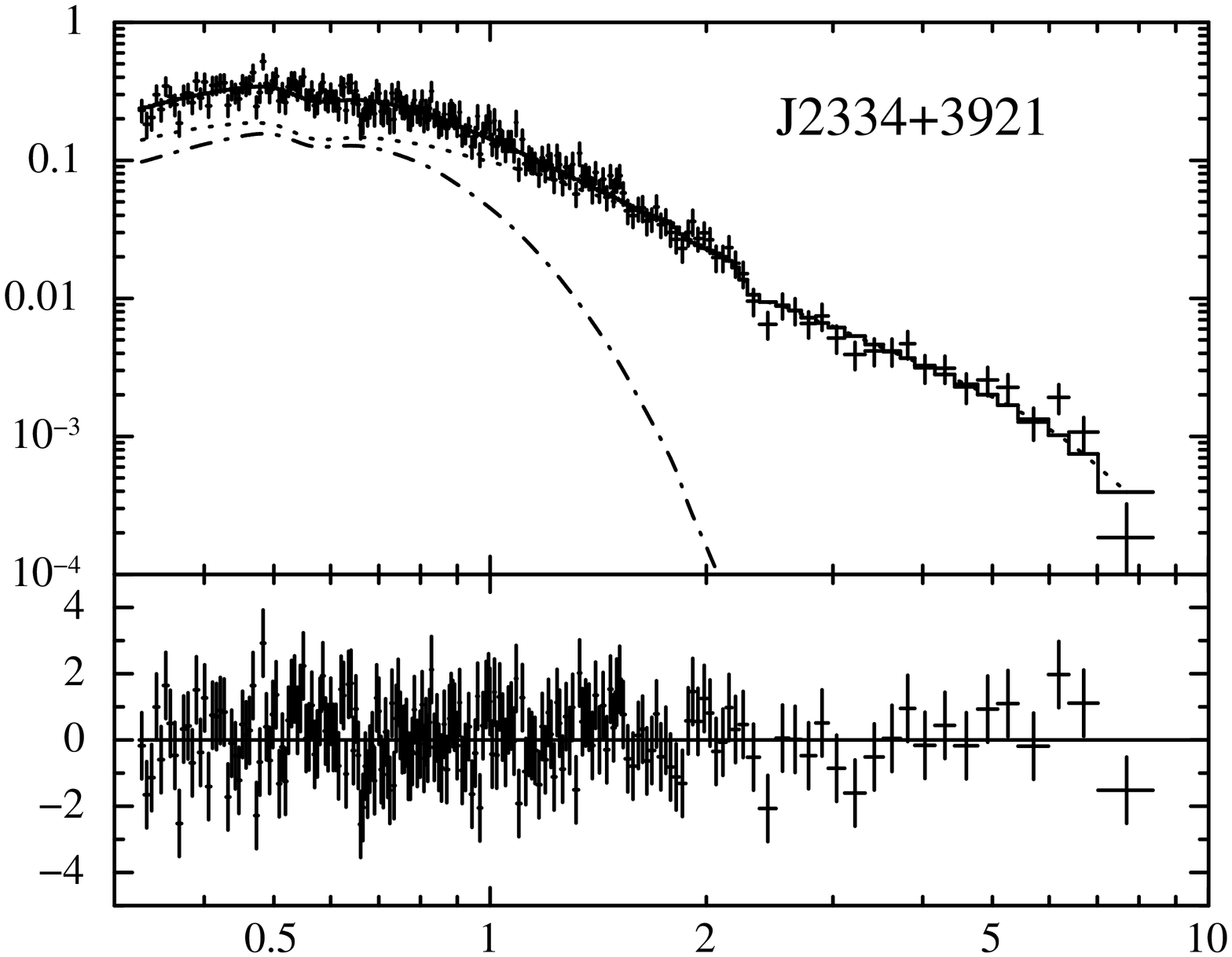}
\includegraphics[width=7.55cm,height=8.75cm,clip,angle=270]{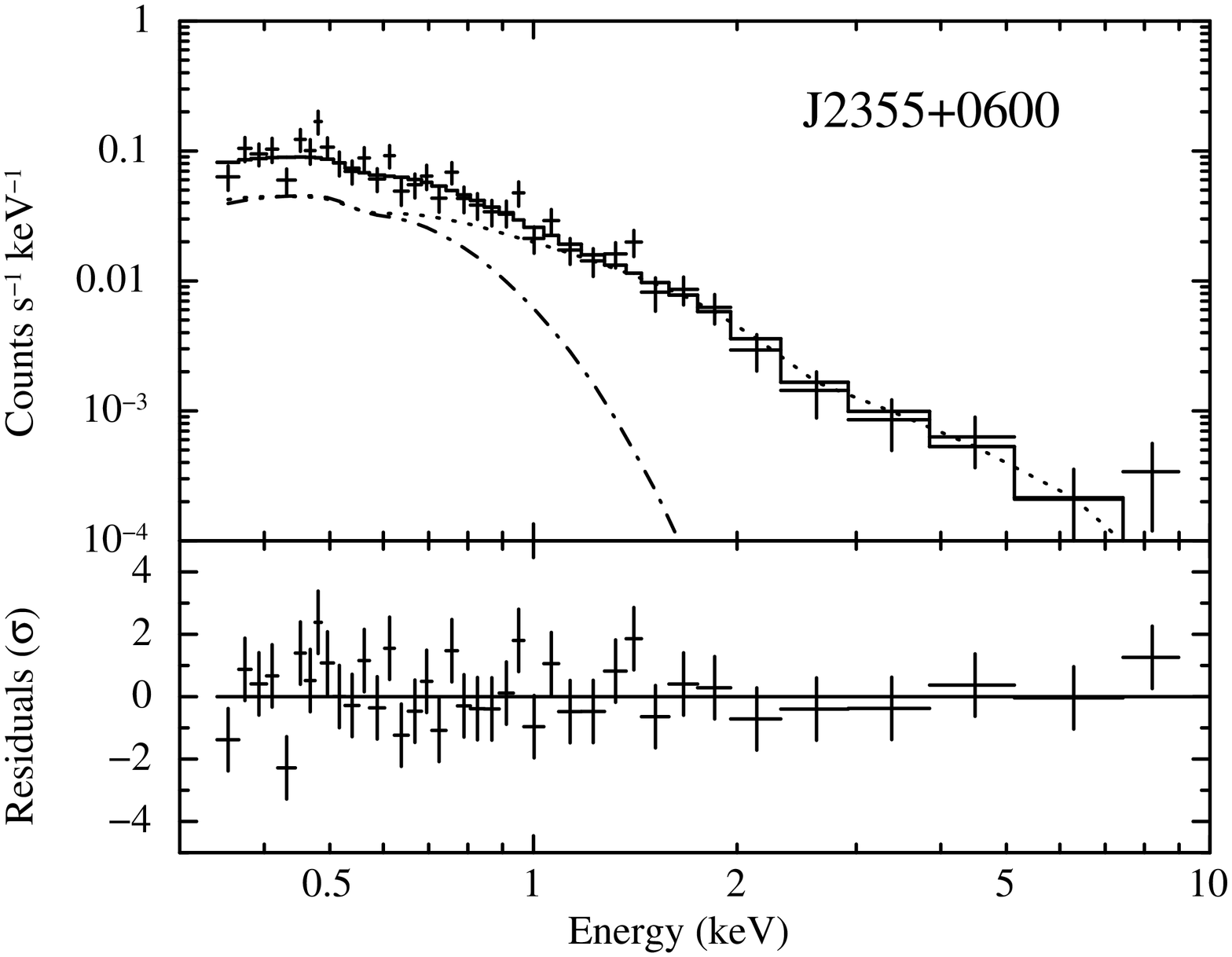}
\caption{Continued}
\end{figure*}


\section{DISCUSSION}

\subsection{Black Hole Mass and Eddington Ratio}

We estimated BH masses based on a correlation between BH mass ($M_{\rm BH}$) and NXS. 
Such correlations have been studied by various authors 
(O'Neill et al. 2005; Papadakis 2004; Miniutti et al. 2009; Niko{\l}ajuk et al. 2009; Zhou et al. 2010). 
These studies used NXS calculated from light curves in the 2$-$10 keV band with 256 s bin. 
We, however, calculated NXS by using light curves in the 0.5$-$10 keV band with 512 s bin.
Then a correlation between BH mass and NXS in 0.5--10 keV
was derived by using a reverberation-mapped sample consisting of 20 AGNs.
These AGNs are the same as used in Zhou et al. (2010).
PG 0026+129 in their sample, however, was excluded from our analysis because no {\it XMM-Newton} data were available.
We obtained
\begin{equation}
M_{\rm BH}=10^{5.76\pm0.13}(\sigma^2_{\rm NXS,0.5-10})^{-0.64\pm0.04}\;M_\odot,\label{eq:mass2}
\end{equation}
where $\sigma^2_{\rm NXS,0.5-10}$ is corrected to the common duration (50 ks) and the common bin size (256 s)
by using the method of Awaki et al. (2006), since NXS depends on the duration and time bin size. 
The energy band we use (0.5-10 keV) is different from that used in the previous studies (2-10 keV).
The 2-10 keV band is likely to be mainly from intrinsic emission unless absorption to the nucleus
is large and variable, and suitable to study the timescales of phenomena just
around BHs. The softer energy band below 2 keV, however,
is more affected by changes in absorber and soft excess emission, of
which the origin is still unknown, and should be treated with caution. Note also that NXS in the soft
X-ray band is strongly correlated with that in the hard X-ray band (Ponti et al. 2012) despite the
different composition of spectral components in the soft and hard bands. This result strongly suggests
that NXS in 0.5-10 keV is well correlated with that in 2-10 keV.

In order to use equation (5) for BH mass estimation,
we corrected NXS for our sample to the common duration (50 ks) and the common bin size (256 s).
Awaki et al. (2006) showed the conversion between
two NXSs ($\sigma^2_{\rm NXS,1}$, $\sigma^2_{\rm NXS,2}$) calculated
with different duration ($T_1,T_2$) and different bin size ($\Delta t_1,\Delta t_2$) as
\begin{equation}
\frac{\sigma^2_{\rm NXS,1}}{\sigma^2_{\rm NXS,2}}=\frac{(2\Delta
t_1)^{\alpha-1}-(T_1)^{\alpha-1}}{(2\Delta t_2)^{\alpha-1}-(T_2)^{\alpha-1}},
\end{equation}
where $\alpha$ is the slope of the PSD, and is assumed to be 2.25.
In Table \ref{table:NXS}, the values of NXS before and after the corrections of duration and bin size,
and estimated BH masses are listed.
The time spans from the first bin to last bin of the light curves, which corresponds
to $T_1$ in equation (6), are also shown in Table \ref{table:NXS}.

\begin{figure}
\includegraphics[width=7.5cm,height=8.5cm,clip,angle=270]{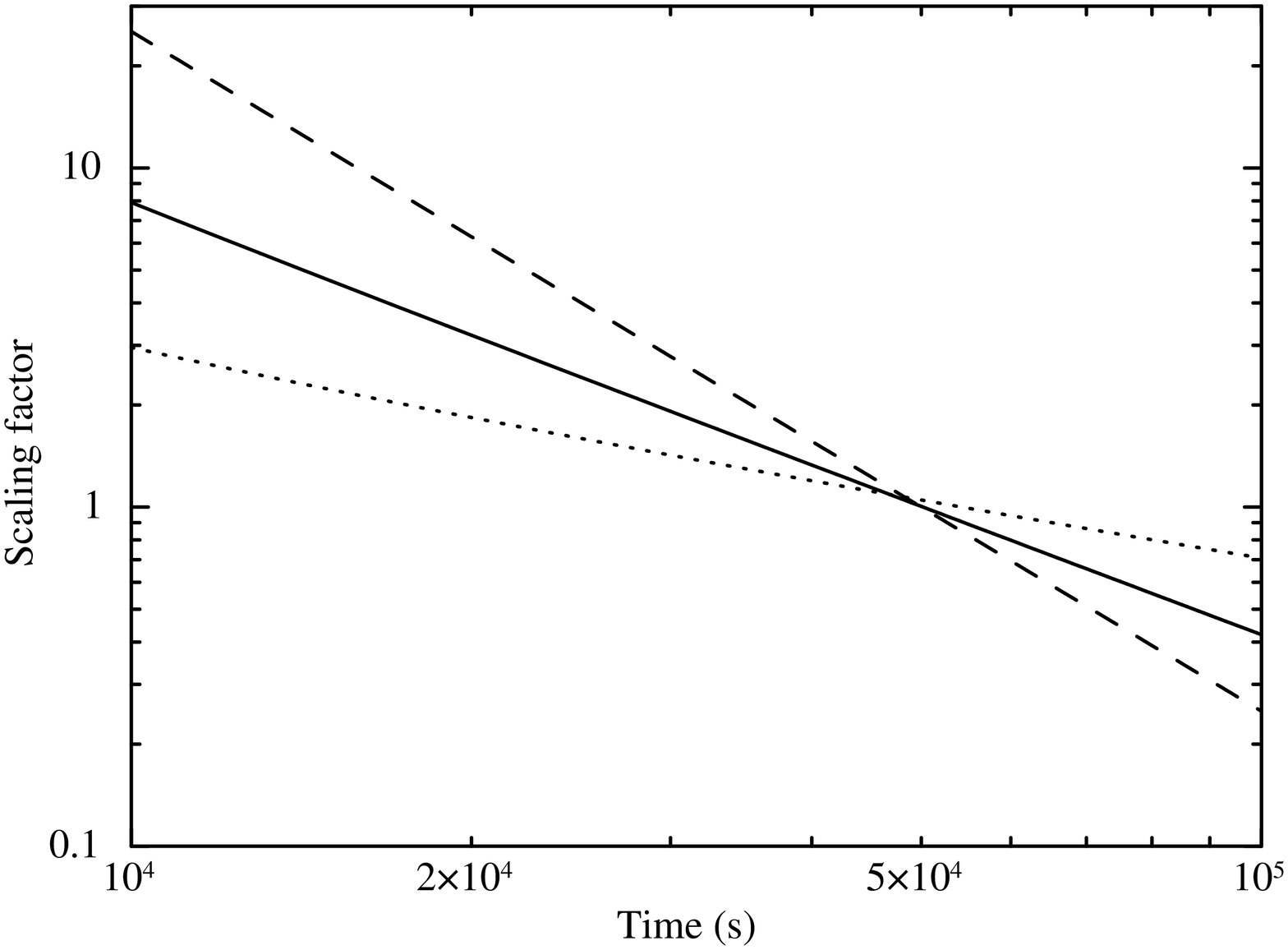}
\caption{Dependence of scaling factor on the duration calculated from equation (6).
Scaling factors for PSD slope of 1.50, 2.25, and 3.00 are shown with dotted, solid, and dashed lines, respectively.}
\label{fig:SF}
\end{figure}

We assumed the PSD slope in 0.5--10 keV to be 2.25, which is typically observed in 2--10 keV
(Papadakis et al. 2002; Uttley, McHardy, \& Papadakis 2002; Edelson \& Nandra 1999;
Markowitz et al. 2003, 2007; Vaughan \& Fabian 2003; McHardy et al. 2004, 2005; Vignali et al. 2004; Awaki et al. 2005;
Markowitz \& Uttley 2005; Uttley \& McHardy 2005).
Although the number of AGNs studied is very limited, PSD slopes in the soft X-ray band appear to be
similar to or only somewhat steeper than those in the 2--10 keV band
(Papadakis et al. 2002; McHardy et al. 2004; Markowitz et al. 2007).
We examined the dependence of a scaling factor of NXS on the duration of the light curve
for the PSD slope of 1.50, 2.25, and 3.00
to estimate an effect of the assumed PSD slope on BH mass estimation.
These slopes were chosen because PSD slopes of most AGNs likely lie within this range.
The scaling factor was calculated from equation (6)
by substituting $T_1$, $\Delta t_1$, and $\Delta t_2$ for 50 ks, 256 s, and 512 s, respectively.
The dependence of the scaling factor on the duration of the light curve is shown in Fig.~\ref{fig:SF}.
The shortest and longest duration in our sample are 14.8 ks and 77.8 ks, respectively.
According to Fig.~\ref{fig:SF}, the influence on NXS by difference of PSD slopes
is larger at 14.8 ks than at 77.8 ks.
The scaling factor for $\alpha=1.50$ and 3.00 divided by those for $\alpha=2.25$ at 14.8 ks are 0.47 and 2.4, respectively.
If NXS multiplied by 0.47 or 2.4 are substituted for $\sigma_{\rm NXS,0.5-10}^2$ in equation (5),
BH masses become smaller or larger by $\sim60$\%, respectively.

The BH masses for our sample were estimated by using equation (5),
and are in the range of $(0.58-6.6)\times10^6\;M_\odot$.
The estimated masses for relatively low-mass BHs could be incorrect due to
the effect of a break in PSD.
NXS represents an integration of the PSD in a certain frequency range
normalized by the mean count rate squared (Vaughan et al. 2003).
If the break frequency is in the frequency range, BH mass based on NXS would be underestimated.
The break time, the reciprocal of the break frequency,
is proportional to BH mass as $M_{\rm BH}/10^{6.5}M_\odot~{\rm day}$ (Markowitz et al. 2003).
In the case of our sample, if the break time is shorter than either 
the duration of the light curve or 50 ks, BH mass would be underestimated.
We calculated the break times of the estimated BH masses
for our sample using the equation in Markowitz et al. (2003).
The break times of seven sources
(J0113$-$1442,
J0136+1549,
J0324$-$0256,
J1201$-$1848,
J1347+1734,
J2008$-$4440, and
J2131$-$4251)
are shorter than 50 ks.
The estimated break times for all the other objects are longer than 50 ks and the length of their light curves.
BH masses of these sources are likely to be underestimated, since NXS are overestimated
due to the scaling using an inappropriate PSD.
We tried to obtain their BH masses taking into account the effect of the break,
by assuming a universal shape of PSD,
a power law of frequency $Af^{-\alpha}$
with $\alpha=0$ and 2.25 at lower and higher frequencies
than the break frequency, where $A$ is a normalization.

First, we calculated the integrals of the PSD with and without the break
from $1/T$ to $1/2\Delta t$, where $T$ and $\Delta t$ are 50 ks and 256 s, respectively.
The integration of the PSD with the break divided by that of the PSD without the break
was calculated.
We multiplied this value by scaled NXS listed in Table~5 to obtain new NXS
and re-estimated BH mass based on new NXS by using equation (5).
The true BH mass should be lower than this re-estimated mass,
since the re-estimated mass is calculated by using the break time for the underestimated mass.
We then found a self consistent solution for a mass, NXS, and a break frequency
inside the mass range determined by the upper and lower bound of mass derived above.
Finally, we obtained the BH masses of the four sources listed in Table~5.
Thus, the BH mass range for our sample becomes $(1.1-6.6)\times10^6\;M_\odot$,
and BH masses of the seven sources are $2\times10^6\;M_\odot$ or less, which are in the range for IMBHs.
If the PSD slope of 1.50 or 3.00 is used instead of 2.25,
the BH masses estimated by these steps are changed by only $1-9$\%.

Growing BHs are expected to have a combination of relatively low-masses and high accretion rates. We calculated Eddington ratios $L_{\rm bol}/L_{\rm Edd}$ for nine objects with known redshifts, where the Eddington luminosity is $1.26\times10^{38}(M_{\rm BH}/M_\odot)$ erg s$^{-1}$. Bolometric luminosities are derived from the intrinsic 2--10 keV luminosities by assuming a bolometric correction factor of 20 (Vasudevan \& Fabian 2007). The Eddington ratios $L_{\rm bol}/L_{\rm Edd}$ thus obtained are listed in Table \ref{table:REDD}.
and the relation between $M_{\rm BH}$ and bolometric luminosities is shown in Fig.$\;$\ref{fig:MBH_L}(a).
Six among nine sources have $L_{\rm bol}/L_{\rm Edd}$ greater than 0.2. This result and the estimated $M_{\rm BH}$ imply that we successfully selected relatively low-mass BHs with a large mass accretion rates, which are likely to be growing BHs.

\begin{deluxetable*}{lcccc}
\tabletypesize{\scriptsize}
\tablecaption{Fluxes, Luminosities, and Eddington Ratios Derived from Best-fit Models\label{table:REDD}}
\tablewidth{0pt}
\tablehead{\colhead{Name} & \colhead{Model$^{\rm a}$} & \colhead{$f_{2-10}$$^{\rm b}$} &\colhead{$L_{2-10}$$^{\rm c}$} & \colhead{$L_{\rm bol}/L_{\rm Edd}$$^{\rm d}$}\\
 & & \colhead{($10^{-14}$ erg cm$^{-2}$ s$^{-1}$)} & \colhead{($10^{42}$ erg s$^{-1}$)} & }
\startdata
J0021$-$1507 & Edge(PL+BB) & 8.7 & 4.1 & 0.26\\
J0113$-$1442 & Edge$\times$PCA$\times$PL & 37 & 3.7 &  0.35\\
& PCA(PL+BB) & 40 & 3.3 & 0.37\\
J0136+1549 & Edge(PL+BB) & 6.5 & \nodata & \nodata\\
J0152$-$1347 & PL+BB & 7.0 & 5.2 & 0.21\\
J0232$-$0729 & PL & 3.3 & 2.4 & 0.12\\
J0324$-$0256 & PL & 15 & \nodata & \nodata\\
J1201$-$1848 & PL+BB & 2.3 & \nodata & \nodata\\
J1233+0005 & PL+BB & 3.1 & 3.4 & 0.08\\
J1305+1813 & PL+BB & 1.2 & 0.99 & 0.07\\
J1324+3000 & PL+BB & 21 & 6.6 & 0.36\\
J1347+1734 & PL+BB & 14 & 0.66 & 0.08\\
J2008$-$4440 & PL+BB & 65 & 5.2 & 0.46\\
J2131$-$4251 & PL+BB & 4.9 & \nodata & \nodata\\
J2334+3921 & PL+BB & 23 & \nodata & \nodata\\
J2355+0600 & PL+BB & 4.6 & \nodata & \nodata\\
\enddata
\tablenotetext{a}{Best-fit model. Edge: absorption edge, PL: power law, BB: blackbody, PCA: partial covered absorption. All components are absorbed by the Galactic column density.}
\tablenotetext{b}{Observed flux in the 2$-$10 keV band.}
\tablenotetext{c}{Intrinsic luminosity in the 2$-$10 keV band.}
\tablenotetext{d}{Eddington ratio.}
\end{deluxetable*}

\begin{deluxetable*}{lccccc}
\tabletypesize{\scriptsize}
\tablecaption{Properties of X-ray Variability\label{table:NXS}}
\tablewidth{0pt}
\tablehead{\colhead{Name} & \colhead{$T_{1}$$^{\rm a}$} & \colhead{$\sigma_{\rm NXS}^2$$^{\rm b}$} & \colhead{scaled $\sigma_{\rm NXS}^2$$^{\rm c}$} & \colhead{$M_{\rm BH}$$^{\rm d}$}
& \colhead{corrected $M_{\rm BH}$$^{\rm e}$}\\
 & \colhead{(ks)} & \colhead{($10^{-3}$)} & \colhead{($10^{-3}$)} & \colhead{($10^6M_\odot$)}
 & \colhead{($10^6M_\odot$)}}
\startdata
J0021$-$1507 & 45.1 & $91\pm30$ & $104\pm34$ & 2.5 & \nodata\\
J0113$-$1442 & 24.6 & $86\pm29$ & $212\pm72$ & 1.6 & 1.7\\
J0136+1549 & 30.2 & $222\pm46$ & $422\pm88$ & 1.0 & 1.4\\
J0152$-$1347 & 50.2 & $50\pm28$ & $50\pm28$ & 3.9 & \nodata\\
J0232$-$0729 & 43.0 & $58\pm30$ & $71\pm37$ & 3.1 & \nodata\\
J0324$-$0256 & 14.8 & $216\pm35$ & $1000\pm160$ & 0.58 & 1.1\\
J1201$-$1848 & 16.9 & $181\pm41$ & $720\pm160$ & 0.71 & 1.2\\
J1233+0005 & 77.8 & $39\pm19$ & $22\pm11$ & 6.6 & \nodata\\
J1305+1813 & 45.1 & $117\pm45$ & $134\pm52$ & 2.1 & \nodata\\
J1324+3000 & 35.8 & $52\pm20$ & $79\pm30$ & 2.9 & \nodata\\
J1347+1734 & 56.3 & $724\pm30$ & $626\pm26$ & 0.78 & 1.3\\
J2008$-$4440 & 39.9 & $141\pm11$ & $188\pm15$ & 1.7 & 1.8\\
J2131$-$4251 & 29.7 & $94\pm30$ & $183\pm57$ & 1.7 & 1.8\\
J2334+3921 & 30.7 & $59.9\pm6.8$ & $111\pm13$ & 2.4 & \nodata\\
J2355+0600 & 33.3 & $82\pm24$ & $138\pm41$ & 2.0 & \nodata\\
\enddata
\tablenotetext{a}{Time span from first bin to last bin of light curve.}
\tablenotetext{b}{Normalized excess variance calculated by using the light curve in the 0.5$-$10 keV band with 512 s bin.}
\tablenotetext{c}{Normalized excess variance after the corrections of duration and bin size.}
\tablenotetext{d}{BH mass estimated in this work by assuming a power law PSD.}
\tablenotetext{e}{BH mass taking into account the effect of the break in PSD (see the text).}
\end{deluxetable*}

\subsection{X-ray Spectra}

\begin{figure}[p]
\includegraphics[width=7.1cm,height=8.5cm,clip,angle=270]{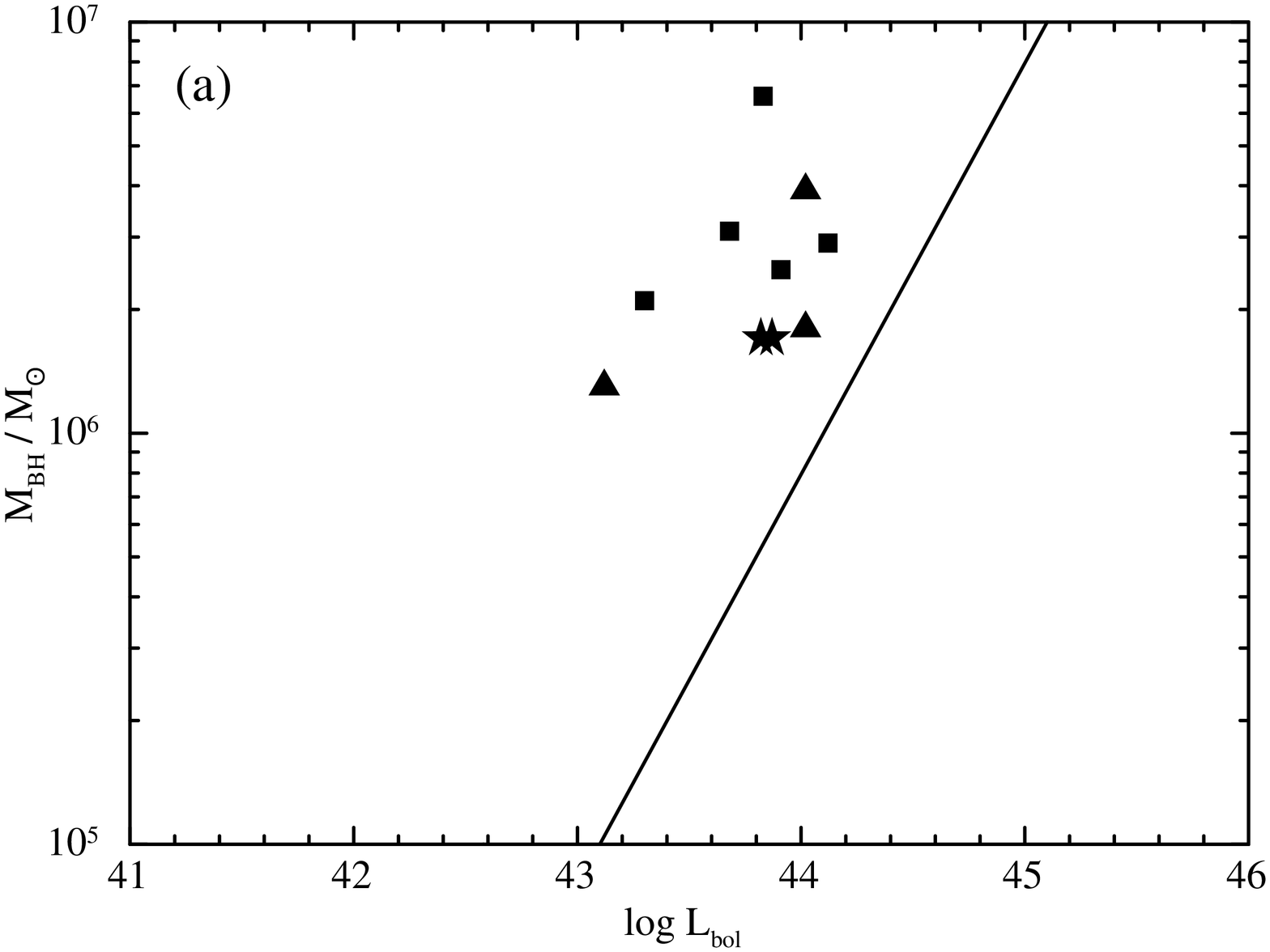}
\includegraphics[width=7.1cm,height=8.5cm,clip,angle=270]{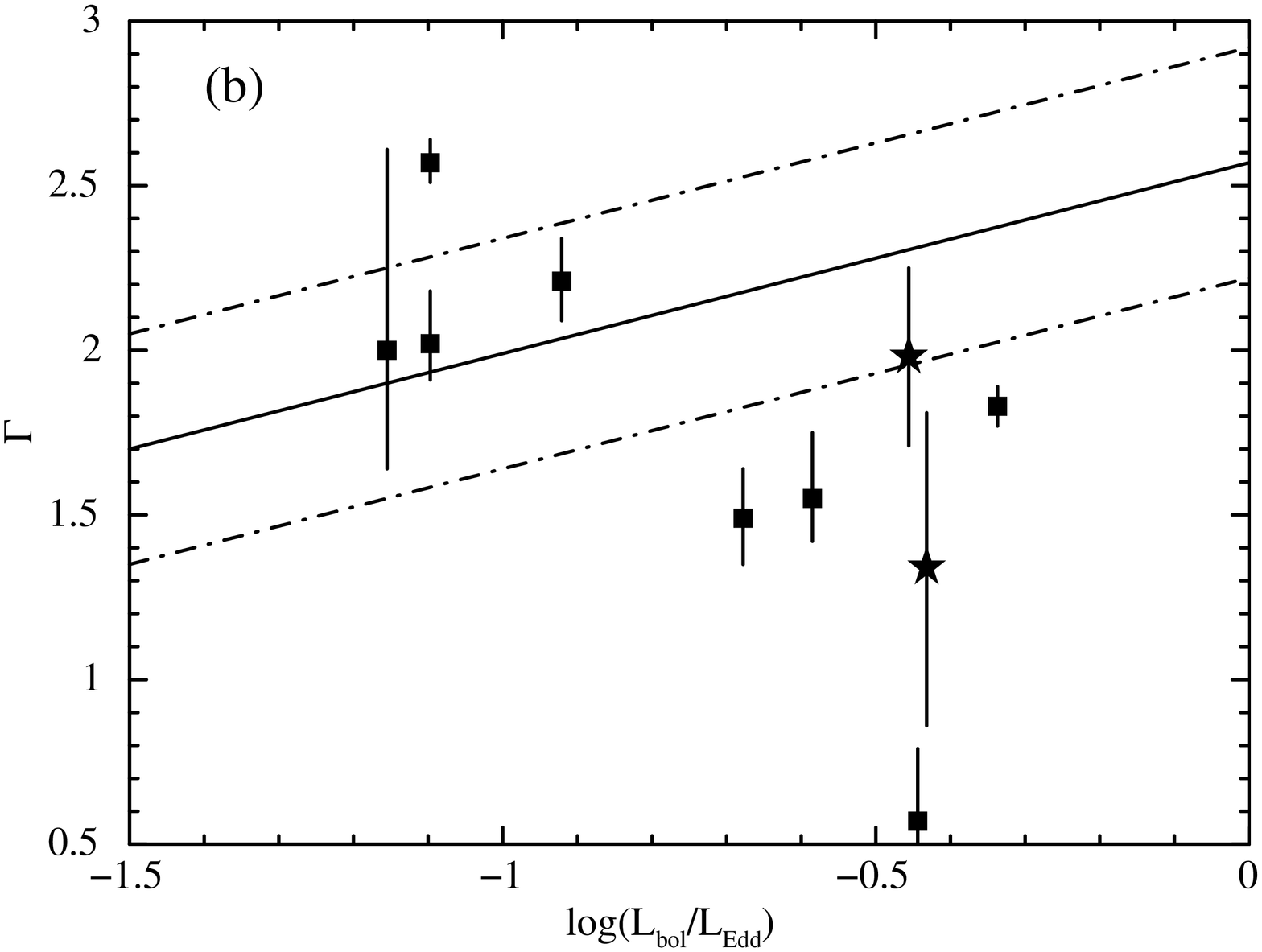}
\includegraphics[width=7.1cm,height=8.5cm,clip,angle=270]{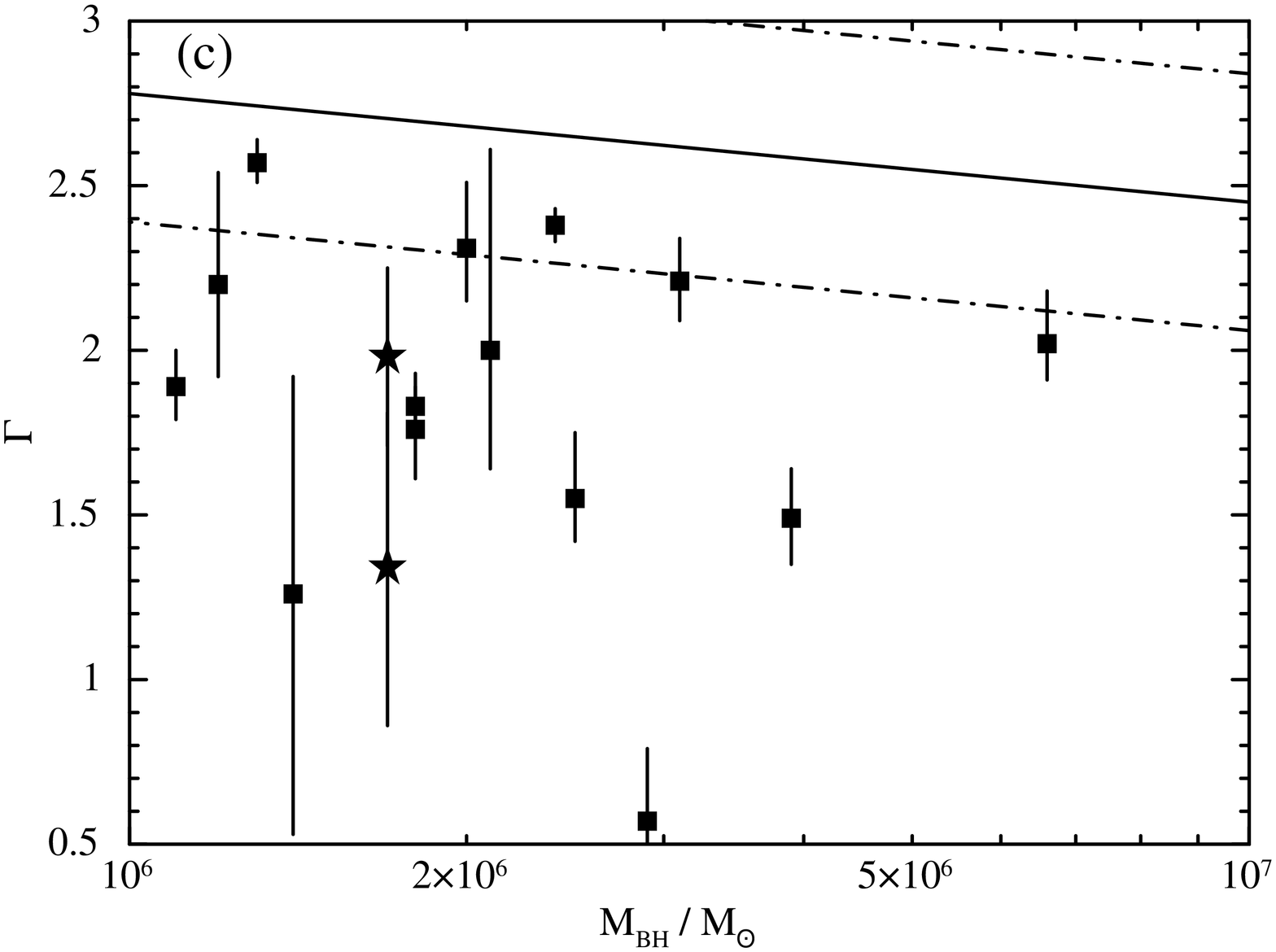}
\caption{(a) BH mass $M_{\rm BH}$ derived from X-ray variability and bolometric luminosity $L_{\rm bol}$.
Luminosities are derived by using spectroscopic redshift (triangles) or photometric redshift (squares).
Solid line corresponds to the Eddington luminosity.
(b) X-ray photon index $\Gamma$ and Eddington ratio $L_{\rm bol}/L_{\rm Edd}$.
Solid line is the relation by Risaliti et al. (2009)
and the dash-dotted lines represent the dispersion.
(c) X-ray photon index $\Gamma$ and BH mass $M_{\rm BH}$ derived
from X-ray variability. Solid and dash-dotted lines correspond
the extrapolation of the relation derived by Risaliti et al. (2009) and the dispersion, respectively.
In all panels, stars denote data points of J0113--1442 for models (3) and (4). Its luminosities are derived by using spectroscopic redshift.}
\label{fig:MBH_L}
\end{figure}

The objects in our sample likely host relatively low-mass SMBHs
accreting at relatively large mass-accretion rates. 
Narrow-line Seyfert 1s (NLS1s) are a class of AGNs, which tend to host relatively low-mass SMBHs with high accretion rates 
(e.g., Czerny et al. 2001; Boroson 2002; Wang \& Netzer 2003) and objects in our sample are expected to share
properties similar to NLS1s.
Many NLS1s show relatively steep X-ray spectra above $\sim2$ keV with photon indices 2.0-2.5, and strong
soft emission relative to underlying power law compared to broad-line Seyfert 1s (BLS1s)
(e.g.,
Boller et al. 1996;
Brandt, Mathur \& Elvis 1997;
Leighly 1999;
Grupe 2004; Bianchi et al. 2009; Caccianiga et al. 2011).
The photon indices observed for 15 sources in our sample are in the range of  $\Gamma\approx0.57-2.57$, which is
much broader than those known for BLS1s and NLS1s.
Photon indices are likely to be related to Eddington ratios; 
a correlation between $\Gamma$ and $L_{\rm bol}/L_{\rm Edd}$ is known for objects with $L_{\rm bol}/L_{\rm Edd}<1$ 
(e.g., Lu \& Yu 1999; Shemmer et al. 2006, 2008; Risaliti et al. 2009). 
We examined a relationship between photon indices $\Gamma$  and Eddington ratios $L_{\rm bol}/L_{\rm Edd}$ 
as shown in Fig.$\;$\ref{fig:MBH_L}(b). 
The relation derived by Risaliti et al. (2009) is also shown as a solid line in Fig.$\;$\ref{fig:MBH_L}(b) 
for comparison. 
The photon indices of about a half of our sources are flatter than expected from Risaliti et al.'s relation.
Ai et al. (2011) suggested that some objects show spectral slopes flatter than the relation 
using a sample of NLS1s with optical broad line widths 
narrower than 1200 km s$^{-1}$ and large Eddington ratios.
We strengthened their finding by adding more data points of objects with low BH masses and large accretion rates
selected by our own method.

One possible reason for such flatness of the slope
could be uncertainties in BH masses.  BH mass estimation relying on X-ray variability is uncertain up to
a dex. The BH masses $M_{\rm BH}$ estimated in the present study may be underestimated, because values of NXS 
from different observations of the same object show nonstationarity and because we tend to select a strongly variable state.
Our comparison between BH masses derived from X-ray variability and from reverberation mapping show
that the discrepancy between the two methods is within a factor of four for most of the objects.
Even if there are uncertainties of $M_{\rm BH}$ by an order of magnitude, our sample contains some objects with flat photon indices
well below Risaliti et al.'s relation. Therefore, uncertainties in BH mass estimation alone cannot explain the
deviation from the relation and the results likely reflect the intrinsic nature of the sources.

We studied also the dependence of photon indices on BH masses. 
The relationship between these quantities is shown in Figure \ref{fig:MBH_L}(c). 
The slopes are again flatter than the extrapolation of the relation of Risaliti et al. (2009). 
As discussed in the previous paragraph, the change of intrinsic slope
could be the reason for the flatness. 
In addition, there could be an uncertainty in Risaliti's relation
because of relatively narrow range of BH masses and Eddington ratios they used. 
Our results along with other X-ray studies of AGNs with IMBHs (Dewangan et al. 2008; Miniutti et al. 2009; Ai et al. 2011) suggest that,
in BH mass range of $10^{4}-10^{6}$ $M_\odot$, photon indices become flatter than the extrapolation of Risaliti et al.'s photon index - BH mass relation.

X-ray spectra of NLS1s, many of which have relatively low-mass SMBHs
with high accretion rates, often show soft X-ray excess over a power-law model 
determined at high energies above 2 keV.
The shape of the soft X-ray excess can be approximated by blackbody or multicolor disk blackbody.
Soft excess emission was also observed in AGNs with IMBHs selected from SDSS (Dewangan et al. 2008). 
When a blackbody model is fitted to the soft X-ray excess, temperature ranges form 80 to 250 eV
for NLS1s and AGNs with IMBHs (Vaughan et al. 1999; Leighly 1999; Dewangan et al. 2008; Ai et al. 2011).
Radio-quiet AGNs having more massive BHs of $10^{7}-10^{9}$ $M_\odot$ show a similar range
of temperature (Gierli\'nski \& Done 2004; Crummy et al. 2006; Bianchi et al. 2009).
In our sample, 12 among the 15 sources showed soft X-ray excess expressed by a blackbody model with $kT\approx83-294$ eV.
Temperatures of all sources expect for J1324+3000 ($kT\approx294$ eV) are consistent with the range
observed for the AGNs mentioned above.
Thus we strengthened the previous results, in which the temperature of soft excess emission
approximated by blackbody is in a relatively narrow range ($80-250$ eV) regardless BH mass.

\vspace{1cm}

\section{CONCLUSION}
We searched for AGNs with relatively low-mass SMBHs (or IMBHs) from 2XMMi-DR3 by using X-ray variability. 
We discovered 16 new AGN candidates showing large amplitude variability. 
NXS of 15 sources expect for 2XMM J123103.2+110648 was
calculated by using light curves in the 0.5$-$10 keV band with 512 s bin. 
NXS were corrected to the common duration (50 ks) and the common bin size (256 s)
by using the method of Awaki et al. (2006) to compare NXS of multiple sources.
BH masses were estimated by using the NXS-$M_{\rm BH}$ derived by our study.
When estimating BH masses, we took into account the effect of the PSD shape.
We obtained large values of NXS and relatively low BH masses in
$(1.1-6.6)\times10^6\;M_\odot$. 
Eddington ratios were calculated for nine sources with known redshifts and found that their accretion rates are high; 
six among them have $L_{\rm bol}/L_{\rm Edd}$ greater than 0.2. 
These results indicate that we successfully selected growing BH candidates by our X-ray variability selection. 

We analyzed X-ray spectra of the 15 objects and compared their X-ray spectral properties with those of AGNs previously
known.
The photon indices $\Gamma$ of about half of our sources are flatter than 
the extrapolation of the trend derived by Risaliti et al. (2009) toward large Eddington ratios, confirming suggestions of Ai et al. (2011).
The soft X-ray excess seen in 12 among 15 sources was expressed by the blackbody model.
The blackbody temperatures for the sources excluding J1324+3000 are in the range of $kT\sim83-167$ eV,
in agreement with the range found in AGNs having BHs of $10^{4}-10^{9}$ $M_\odot$.

\acknowledgements

We thank the anonymous referee for helpful comments and suggestions that improved the paper.
This work was supported by the Grant-in-aid for Scientific Research
20740109 (YT) from the Ministry of Education, Culture, Sports, Science,
and Technology of Japan.

{\it Facility:} XMM

\end{document}